\documentclass[fleqn,usenatbib]{mnras}

\usepackage{newtxtext,newtxmath}

\usepackage[T1]{fontenc}
\usepackage{ae,aecompl}

\usepackage{graphicx}
\usepackage{longtable}
\usepackage[labelfont=bf]{caption}
\usepackage{lscape,supertabular}
\usepackage{natbib}
\usepackage{hyperref}
\usepackage{subfig}
\usepackage{enumitem}
\usepackage{inputenc}
\usepackage{multirow}
\usepackage[range-units = brackets,tophrase={-},seperr,repeatunits=false]{siunitx}

\title[HICOSMO II]{HICOSMO - Cosmology with a complete sample of galaxy clusters. \\\Large II. Cosmological results}

\author[Schellenberger et al.]{
G. Schellenberger$^{1, 2}$\thanks{E-mail: gerrit.schellenberger@cfa.harvard.edu},
and T. H. Reiprich$^{1}$
\\
$^{1}$Argelander-Institut f\"ur Astronomie, Universit\"at Bonn, Auf dem H\"ugel 71, 53121 Bonn, Germany\\
$^{2}$Harvard-Smithsonian Center for Astrophysics, 60 Garden Street, Cambridge, MA 02138, USA
}

\date{Accepted XXX. Received YYY; in original form ZZZ}

\pubyear{2017}

\begin{document}
\label{firstpage}
\pagerange{\pageref{firstpage}--\pageref{lastpage}}
\maketitle

\begin{abstract}
The X-ray bright, hot gas in the potential well of a galaxy cluster enables systematic X-ray studies of samples of galaxy clusters to constrain cosmological parameters.
HIFLUGCS consists of the 64 X-ray brightest galaxy clusters in the Universe, building up a local sample. Here we utilize this sample to determine, for the first time, individual hydrostatic mass estimates for all the clusters of the sample and, by making use of the completeness of the sample, we quantify constraints on the two interesting cosmological parameters, $\Omega_\mathrm{m}$ and $\sigma_8$.
We apply our total hydrostatic and gas mass estimates from the X-ray analysis to a Bayesian cosmological likelihood analysis and leave several parameters free to be constrained.
We find $\Omega_\mathrm{m} = \num{0.30(1)}$ and $\sigma_8 = \num{0.79(3)}$ (statistical uncertainties, 68\% credibility level) using our default analysis strategy combining both, a mass function analysis and the gas mass fraction results.
The main sources of biases that we correct here are (1) the influence of galaxy groups (incompleteness in parent samples and differing behavior of the $L_x-M$ relation), (2) the hydrostatic mass bias, (3) the extrapolation of the total mass (comparing various methods), (4) the theoretical halo mass function and (5) other physical effects (non-negligible neutrino mass). We find that galaxy groups introduce a strong bias, since their number density seems to be over predicted by the halo mass function. On the other hand, incorporating baryonic effects does not result in a significant change in the constraints. The total (uncorrected) systematic uncertainties ($\sim 20\%$) clearly dominate the statistical uncertainties on cosmological parameters for our sample.	
\end{abstract}



\section{Introduction}
\label{sec:intro}
Galaxy clusters reside at the intersections of Dark Matter filamentary structure. These largest gravitationally bound systems therefore bear witness to  the growth of structure in the Universe and are excellent objects for cosmological studies. With our current knowledge of cosmological parameters it is possible to construct the cluster mass function and predict the number density of Dark Matter halos, reflected by the observed population of galaxy clusters. The key parameters are the normalized matter density, $\Omega_{\rm M}$, and the amplitude of initial density fluctuations, $\sigma_8$. Even using only nearby galaxy clusters these quantities can be constrained from the shape of the mass function (e.g., \citealp{Ikebe2002,reiprich_hiflugcs}).

In the past, tremendous efforts were made to break the degeneracy between these parameters and lower their uncertainties by using better instruments, larger samples of clusters, and more advanced analysis methods.

However, there are alternative approaches to constraining cosmology which can be applied alongside or in combination with studies of galaxy clusters. The primary anisotropies in the cosmic microwave background (CMB) are an independent probe of cosmological parameters, which trace the radiation from a very early epoch of the Universe. This makes them a complementary tool to galaxy clusters, which trace structure formation across cosmic time.
The ellipses of the $\Omega_{\rm M}$ - $\sigma_8$ confidence levels are almost perpendicular to each other, which means including different probes in the analysis will strongly shrink the uncertainties.
Although the combined analysis has advantages, it is crucial to interpret the cosmological results of each method separately to estimate and minimize systematic biases. 

Assumptions have to be made to obtain the galaxy cluster total mass from X-ray observations. One way is to assume that the ICM is in hydrostatic equilibrium. Another is to use tracers like luminosity or temperature as a proxy for the total mass. 
Calibrating the scaling relations of between these observables and the total mass using, e.g., weak gravitational lensing observations, might provide a way to be  avoid biases caused by deviations from hydrostatic equilibrium. Unfortunately other (maybe unknown) biases are connected with weak lensing studies, such as noise bias (e.g., \citealp{2013MNRAS.429..661M}), mass sheet degeneracy (e.g., \citealp{1995A&A...294..411S,2004A&A...424...13B}), asymmetry of the point spread function (e.g., \citealp{2003MNRAS.343..459H}), false photometric redshifts and miscentering (\citealp{2015arXiv150805308K}). Selection effects in the generation of a galaxy cluster sample (e.g., selecting only massive or intrinsically brighter objects) can also bias cosmological results, if they are not accounted for properly.

Moreover, all measurements depend on the calibration accuracy of the instrument used. Any systematic uncertainties arising from the instrument itself have to be known. As shown, e.g., in \citet{Schellenberger2015}, instrumental calibrations are still relatively uncertain in the X-ray regime. Although the relative differences between two instruments, Chandra ACIS and XMM-Newton EPIC, are known, it cannot be determined which instrument is correct, if any. The known uncertainties between these two X-ray instruments can be incorporated as a range of instrumental systematics, but the true cluster temperature could be outside the range given by these two instruments, even though it has been shown with small samples that Chandra-ACIS and XMM-Newton PN mark roughly the extreme cases among 10 X-ray detectors (\citealp{2013arXiv1305.4480G,2014arXiv1412.6233B}). However, as described in \citet{Schellenberger2015}, the impact of the cross calibration uncertainties between XMM-Newton and Chandra on cosmology is not larger than the statistical uncertainties for samples like HIFLUGCS, because most clusters have relatively low temperatures at $R_{500}$. At these outer radii the temperature has dropped significantly for most clusters and groups compared to the peak temperature at inner radii (e.g., \citealp{DeGrandi2002,2005ApJ...628..655V,2007MNRAS.380.1554R,2008A&A...486..359L,2009ApJ...693.1142S}). 

An X-ray flux limited sample like HIFLUGCS is of special interest for cosmology: It provides high quality data of nearby galaxy clusters, which can be studied in detail to get precise temperature and surface brightness profiles. It has been shown in \citet{reiprich_hiflugcs} that with such a sample, $\Omega_{\rm m}$ and $\sigma_8$ can be quantified, so one has an independent probe for cosmological parameters in hand.

In the present study we aim to put constraints on at least these two cosmological parameters and also to gain knowledge about the physical processes in the X-ray brightest galaxy cluster sample.
For the first time we will use individual, X-ray derived hydrostatic mass estimates of a complete sample of galaxy clusters to constrain cosmological parameters. The HIFLUGCS sample consists of the X-ray brightest galaxy clusters, with very high data quality available. Not only will we study the halo mass function and the cosmological implications, but also evaluate and quantify many sources of systematic biases.

In \cite{2017arXiv170505842S} (Paper I), we described in detail our data analysis strategy and presented the total and gas mass estimates for all HIFLUGCS clusters. 
Here we apply these data in a cosmological likelihood analysis and discuss in detail the systematic effects entering in such a flux limited sample of galaxy clusters.
All uncertainties are 68.3\% percent levels, unless stated otherwise. For the calculation of physical cluster quantities we assume a Flat $\Lambda$CDM cosmology with $H_0 = \SI{71}{km\,s^{-1}\,Mpc^{-1}}$ and $\Omega_\mathrm{m} = 0.27$. 

\section{Cosmological analysis}
We determined the important quantities, gas mass and total mass, for each HIFLUGCS galaxy cluster individually. In this section we describe how we constrain cosmological parameters from these two quantities. The model for the total mass is based on the halo mass function and described in following section (\ref{ch:likelihood}). 
The gas mass fraction analysis (Section \ref{ch:gasmass_result}) is based on the idea that the gas mass to total mass fraction should reach a cosmic mean value in galaxy clusters at large radii.

\subsection{Hydrostatic masses}
In Paper I, we describe in detail the derivation of the hydrostatic masses for all HIFLUGCS clusters from Chandra data. We summarize here the important steps:
\begin{itemize}
	\item We parameterize the deprojected cluster temperature and density profiles. For the temperature profile, we use three different models depending on the available data quality, whereas for the surface brightness analysis we use a double-$\beta$ model to determine the density. The covariance between the free parameters is taken into account by an MCMC fitting algorithm.
	\item Assuming hydrostatic equilibrium we can calculate a total mass within a given radius from the MCMC chains. To reach an overdensity of 500 ($R_{500}$) we have to extrapolate for most of the clusters. After comparing several methods, we decided to use the \textit{NFW Freeze} method as default: An NFW model is fitted to the mass profile at a radial range that corresponds to the last three to five temperature measurements. The concentration parameter of the NFW model is set by the $c-M$ relation given in \cite{2013ApJ...766...32B}.
	\item A comparison of the masses with dynamical and SZ masses results in overall agreement.
	\item A small correction for the extrapolation algorithm is applied before they are used in our cosmological analysis. This correction is discussed in Section \ref{ch:r500}.
\end{itemize}

\subsection{The likelihood estimate for the halo mass function}
\label{ch:likelihood}
The halo mass function is sensitive to cosmological parameters like $\Omega_\mathrm{m}$ and $\sigma_8$. It represents the galaxy cluster number density at a given mass and redshift. In order to account for the selection criterion, luminosities for each cluster are required. More details on the input parameters (total cluster mass, redshift and luminosity) are given in Paper I.
In order to construct a cluster mass function, the selection function of the sample is crucial. In the case of HIFLUGCS it is based on an X-ray flux cut. Since there exists a tight correlation between the cluster total mass and the luminosity, we include this scaling relation in the mass function analysis to correct for selection effects (see Appendix \ref{ap:likelihood} for details). A simultaneous fit by leaving scaling relation parameters free to vary ensures that the $L_x-M$ relation is not affected by biases (e.g., Malmquist or Eddington bias), but reflects the behavior of the real cluster population given by the halo mass function.

As shown in Appendix \ref{ap:likelihood} the likelihood function, which evaluates the probability of observed cluster properties based on model parameters, is given by,
\begin{equation}
\label{eq:likelihood21}
\mathcal{L} \propto  e^{- \langle N_\mathrm{det} \rangle} \prod \limits_{i=1}^{N_\mathrm{det}} \langle N_\mathrm{det} \rangle \cdot \tilde{P}_{i}~,
\end{equation}
as also shown in  \citet{2010MNRAS.406.1759M,Mantz2015}.

The implementation in a \verb|C| code also includes the CLASS source code (\citealp{2011arXiv1104.2932L}) to recalculate the transfer function for each cosmology.
For the cosmological application the Metropolis algorithm was implemented. In the following we describe the free parameters and their priors:
\begin{itemize}
	\item $\Omega_\mathrm{m}$, the normalized matter density of today's Universe. It is set to a (flat) uniform probability distribution as prior with 0.05 and 0.5 as the lower and upper limits.
	\item $\sigma_8$, the amplitude of density fluctuations in the initial density field: Uniform prior with 0.3 and 1.2 as boundaries.
	\item $(1-b)$, the mass bias: Can be frozen to 1 (no bias, as in case of the ``raw'' analysis procedure), but in some cases variable with a uniform or normal prior (see Section \ref{ch:disc_gcphysics}).
\end{itemize}
For the parameters of the mass($M$)-luminosity($L_x$, in the $0.1-2.4\,$keV band) relation,
\begin{equation}
\label{eq:lxm}
\log_{10} \left( \frac{L_x}{h^{-2}\,\SI{e44}{erg\,s^{-1}}} \right) = A_\mathrm{LM} + B_\mathrm{LM} \cdot \log_{10} \left( \frac{M}{h^{-1}\,\SI{e15}{M_\odot}} \right):
\end{equation}
\begin{itemize}
	\item $A_\mathrm{LM}$, the intercept of the $L_x-M$ relation (see Eq. \ref{eq:lxm}): Uniform prior with 0 and 2.8 as boundaries.
	\item $B_\mathrm{LM}$, the slope of the $L_x-M$ relation (see Eq. \ref{eq:lxm}): Uniform prior with 0.8 and 2.5 as boundaries.
	\item $\sigma_\mathrm{LM}$, the scatter of the $L_x-M$ relation: By defaults frozen to the scatter of the observed sample, but in some cases variable between 0.1 and 0.5 with a uniform prior.
\end{itemize}
The following parameters were not variable during the cosmological analysis, because either the local halo mass function is not sensitive to them, or they are highly degenerate with other parameters for any galaxy cluster sample and need supplementary methods (e.g., primary CMB anisotropies) to be determined:
\begin{itemize}
	\item $\Omega_\mathrm{k}$, the spatial curvature parameter. It is set to 0, which implies a flat Universe.
	\item $\Omega_\mathrm{r}$, the radiation density of today's Universe, which is set to 0 for all calculations except the transfer function.
	\item $w$, the equation of state parameter for the Dark Energy. This is set to a constant value (no evolution term $w_a$) of $-1$ (i.e. equivalent to a Cosmological Constant).
	\item $N_\mathrm{eff}$, the effective number of neutrino species. This is by default set to $\num{3.046}$ (\citealp{1982PhRvD..26.2694D,2002PhLB..534....8M,2006PhR...429..307L}). Usually one massive neutrino with $m_\nu = \SI{0.06}{eV}$ enters in the calculation of the transfer function (\citealp{2012arXiv1212.6154L,2014NJPh...16f5002L}).
	\item $T_0$, the CMB temperature. It is set to $\SI{2.72548}{K}$ (\citealp{2009ApJ...707..916F}). For the given parameters this means that the photon energy density $\Omega_\gamma = \num{5.0e-5}$ and the neutrino density $\Omega_\nu = \num{3.5e-5}$, so the total radiation density used for the transfer function is $\Omega_\mathrm{r} = \num{8.5e-5}$. More details are also given in Section \ref{ch:neutrinos}.
\end{itemize}
The values are adopted from WMAP9 data (\citealp{Hinshaw2013}), unless stated otherwise. For a discussion of the differences between WMAP and Planck results, and why we rely on the former, we refer to Section \ref{ch:diff_wmap9}. For the Hubble constant, $H_0 = \SI{100}{km\,s^{-1}\,Mpc^{-1}}\cdot h$, the Baryon density $\Omega_\mathrm{b}$, and the spectral slope of the primordial power spectrum $n_s$, the WMAP9 posterior distribution was adopted as a prior\footnote{These priors were estimated from the WMAP9 parameter posterior distribution by a multivariate Gaussian.} on these parameters.

\subsection{Gas mass fraction as cosmological probe}
\label{ch:gasmass_result}
It has been shown that the cluster ICM mass (called gas mass) can be extracted from X-ray observations. Using both the total and gas mass, one can directly draw conclusions on the baryon fraction in the Universe (e.g., \citealp{1993Natur.366..429W}):
\begin{equation}
\label{eq:baryonfrac}
\frac{M_\mathrm{gas} + M_\mathrm{stars}}{M_\mathrm{tot}} \approx \frac{\Omega_\mathrm{b}}{\Omega_\mathrm{m}}~,
\end{equation}
where $M_\mathrm{stars}$ is the total stellar mass, which provides an additional minor contribution to the baryon budget. Observations (e.g., \citealp{2003MNRAS.344L..13E,2003A&A...398..879E}) have shown that there exists a baryon deficit in clusters, which may be interpreted as undetected baryons or underestimated $\Omega_\mathrm{m}$. 
However, equation \ref{eq:baryonfrac} only holds if the gas is distributed in clusters in the same way as in the rest of the Universe. Furthermore, the radius at which the masses are measured should not change the gas mass fraction $f_\mathrm{gas}$. In reality processes like AGN feedback and star formation can cause the gas to be ejected from potential wells. These effects are more dominant in galaxy groups than in clusters, so there may be a mass dependence of the potential baryon deficit.

Non-radiative simulations, including gravity, pressure gradients and hydrodynamical shocks by \citet{1998ApJ...503..569E} or \citet{2007MNRAS.377...41C} predict that the baryon fraction within the virial radius of galaxy clusters is equal or close to the cosmic mean value inferred from CMB experiments. At smaller radii (e.g., $r_{2500}$) a depletion factor can be calculated. These correction factors have been used for cosmology, e.g., by \citealp{2002MNRAS.334L..11A,2008MNRAS.383..879A}.
Other physical effects like AGN feedback, can introduce an additional bias, which has been implemented in more realistic simulations, e.g., by \citet{2013MNRAS.431.1487P} and \citet{2013ApJ...777..123B}, and used, e.g., by \citet{2014MNRAS.440.2077M}, who find $\Omega_\mathrm{m} = \num{0.29(4)}$ and consistency with a flat Universe.

From the CMB power spectrum one has a very precise measurement of the baryon density, $\Omega_\mathrm{b} = \num{0.02222}\,h^{-2}$ (\citealp{2015arXiv150201589P}). \citet{2013MNRAS.431.1487P} provide estimates for the gas mass fraction of galaxy clusters from simulations, 
\begin{equation}
\label{eq:fgasmodel}
f_\mathrm{gas}^{\Lambda\mathrm{CDM}} = Y_\mathrm{g} \,A^{\Lambda\mathrm{CDM}} \, \frac{\Omega_\mathrm{b}}{\Omega_\mathrm{m}} \left( \frac{d_\mathrm{A}^{\Lambda\mathrm{CDM}}(z)}{d_\mathrm{A}(z)} \right)^{1.5}~,
\end{equation}
where $\Lambda$CDM refers to the reference cosmology used for the gas and total mass calculation (flat $\Lambda$CDM model with $h = 0.71$ and $\Omega_\mathrm{m}=0.27$). 
$Y_\mathrm{g}$ is the gas depletion factor. \citet{2013MNRAS.431.1487P} used the WMAP7 (\citealp{2011ApJS..192...18K}) cosmology, which simplifies to $\frac{\Omega_\mathrm{b}}{\Omega_\mathrm{m}} = 0.167$. 
We use the authors model of the gas depletion factor depending on radius, redshift and mass,
\begin{equation}
\label{eq:fgasmodel2}
Y_\mathrm{g} = Y_{0,\mathrm{g}} (1+\alpha z)\,\left(\frac{M_{500}}{\SI{5e14}{h^{-1}\,M_\odot}}\right)^{\beta_\mathrm{gas}} \,\left(\frac{\Delta}{500}\right)^\gamma~,
\end{equation}
where $Y_{0,\mathrm{g}}$ is the normalization, $\alpha$ the redshift dependence, $\beta_\mathrm{gas}$ the mass dependence and $\gamma$ the radial dependence. Priors on these parameters are adopted from their simulations (see Tab. \ref{tab:fgas1}). 
The ratio of angular diameter distances reflects the mass dependence on cosmology\footnote{The power of 1.5 is due to the dependence $M_\mathrm{gas} \propto h^{-\frac{5}{2}}$ and $M_\mathrm{tot} \propto h^{-1}$.}.
$A^{\Lambda\mathrm{CDM}}$ is a correction factor to account for the fact that an overdensity rather than a fixed radius is used to compute the masses, which changes according to cosmology,
\begin{equation}
A^{\Lambda\mathrm{CDM}} = \left( \frac{\theta_{2500}^{\Lambda\mathrm{CDM}}}{\theta_{2500}} \right)^\eta \approx \left( \frac{H(z)\,d_\mathrm{A}}{[H(z)\,d_\mathrm{A}]^{\Lambda\mathrm{CDM}} } \right)^\eta~.
\end{equation}
We used $\eta=0.71$ from our measurements of the slope of the $f_\mathrm{gas}$ profile. \citet{2008MNRAS.383..879A} used $\eta = \num{0.214(22)}$, which does not change our results since the influence of $A^{\Lambda\mathrm{CDM}}$ is very small.

We chose to measure the gas mass fraction at overdensity $\Delta=2500$ (in contrast to the mass function analysis, which uses $\Delta=500$), which only requires an extrapolation of profiles for less than half of the 64 clusters. Furthermore, in this context the total mass is measured from direct temperature profile extrapolation (``kT extrapolate''), to minimize the model influence which might enter with an NFW model. 

We also include recent results on the hydrostatic bias by \citet{2016ApJ...827..112B} into our model. 
At an overdensity of $\Delta=2500$ (unlike at $\Delta=500$) there is a significant difference between the bias for cool core (CC) and non-cool core clusters: We adopt the Gaussian distributed priors $(1-b)_\mathrm{CC} = \num{0.999(27)}$ and $(1-b)_\mathrm{NCC} = \num{0.877(11)}$ and use the classification of CC and NCC clusters by \citet{hudson_what_2009}. Assuming the average gas density profile follows, in the outer regions, a $\beta$-model with $\beta=\frac{2}{3}$, it can be shown that the gas mass increases due to the increase of the radius as:
\begin{equation}
	M_\mathrm{gas} \propto \int \frac{x^2}{1 + x^2} \,\mathrm{d}x~ \approx x - \pi/2,
\end{equation}
where $x = \frac{r}{r_c}\gg 1$, and $r_c$ is the core radius. One can then derive the following dependence for the bias corrected gas mass,
\begin{equation}
	M_\mathrm{gas}^\mathrm{BC} = M_\mathrm{gas} \cdot (1-b)^{-\frac{1}{3}}~,
\end{equation}
and for the gas mass fraction,
\begin{equation}
	f_\mathrm{gas}^\mathrm{BC} = f_\mathrm{gas} \cdot (1-b)^{\frac{2}{3}}~.
\end{equation}
The fit was performed using an MCMC with priors on 8 variables and leaving only $\Omega_\mathrm{m}$ with a uniform prior (see Tab. \ref{tab:fgas1}).
\clearpage
\section{Results}
Our cosmological results from the halo mass function and the gas mass fraction analysis are presented here. Since the $f_\mathrm{gas}$ test only constrains $\Omega_\mathrm{m}$, we present these results first, in order to add these results as priors to our mass function analysis and obtain combined results.

The posterior results in Table \ref{tab:fgas1} for the sample with mass cut (comprising 28 objects), shows excellent agreement with the priors, while $\Omega_\mathrm{m}$ is significantly higher than in the halo mass function  analysis. 
Without the redshift cut $\Omega_\mathrm{m}$ is 14\% lower and the mass dependence of the depletion, $\beta_\mathrm{gas}$, is in strong tension with the prior input (from simulations by \citealp{2013MNRAS.431.1487P}), but in rough agreement with other studies
(\citealp{2009ApJ...693.1142S,2011A&A...535A..78Z}: $\beta_\mathrm{gas} = \num{0.30(7)}$, also \citealp{2006ApJ...640..691V,2007A&A...474L..37A,2009ApJ...693.1142S}: $\beta_\mathrm{gas}=\num{0.21(3)}$ and \citealp{2016MNRAS.455..258C}: $\beta_\mathrm{gas}=\num{0.22(6)}$), especially considering our broad mass coverage.
There exist correlations (absolute of Pearson coefficient larger than 0.5) for $\Omega_\mathrm{m}$ and $h$, $\Omega_\mathrm{m}$ and $\gamma$, and $\Omega_\mathrm{m}$ and $Y_{0,\mathrm{g}}$. The first correlation is negative (the larger $\Omega_\mathrm{m}$, the smaller becomes $h$), while the remainder are positive. We also tested the algorithm by artificially multiplying all $f_\mathrm{gas}$ values by 2 or 0.5. This leads to roughly halved or doubled $\Omega_\mathrm{m}$ values, respectively, while the change for the other parameters is very small. This test was performed both with and without the mass cut, and indicates that the best-fit determination of $\Omega_\mathrm{m}$ is not biased by the priors on the other parameters.
As discussed in Section \ref{ch:groups}, to exclude the effects of biases arising from the structure of the local Universe, we use the redshift limited results ($z>0.05$) as our default. This analysis leads to $\Omega_\mathrm{m} = \num{0.305(9)}$.
\begin{table}
	\renewcommand{\arraystretch}{1.5}
	\small
	\centering
	\begin{tabular}{cccc}
		\hline
		Parameter & Prior & \multicolumn{2}{c}{Posterior} \\
		\hline
		& &  $z>0.05$  &  All clusters \\
		\hline
		$\Omega_\mathrm{m}$ & $U(0.05,0.8)$ & $\num{0.305(9)}$ &  $\num{ 0.260(8)}$\\
		$h$ & $N(0.70,0.022)$ & $\num{0.689(23)}$ &  $\num{ 0.685(22)}$ \\
		$Y_{0,\mathrm{g}}$ & $N(0.67,0.01)$ & $\num{ 0.675(10)}$ &  $\num{0.676(9)}$\\ 
		$\alpha$ & $N(0.02,0.02)$ &   $\num{0.046(2)}$ &  $\num{0.059(2)}$\\
		$\beta_\mathrm{gas}$ & $N(0.06,0.01)$ &  $\num{0.031(7)}$ &  $\num{0.343(2)}$\\
		$\gamma$ & $N(-0.12,0.01)$ &   $\num{-0.115(10)}$ &  $\num{-0.115(10)}$\\
		$(1-b)_\mathrm{CC}$ & $N(0.999,0.027)$ & $\num{1.025(16)}$ &  $\num{1.160(15)}$\\
		$(1-b)_\mathrm{NCC}$ & $N(0.877,0.011)$ &$\num{0.869(10)}$ &  $\num{0.836(10)}$\\
		\hline
		\hline
	\end{tabular}
	\caption{8 free parameters (with priors) for the $f_\mathrm{gas}$ test. $N(x,y)$ is a normal distributed prior with mean $x$ and standard deviation $y$, while $U(v,w)$ is a uniform distributed prior with $v$ and $w$ as the lower and upper boundary.}
	\label{tab:fgas1}
\end{table}

Starting from the 64 HIFLUGCS $M_{500}$ masses which are extrapolated using an NFW model with a frozen concentration parameter (``NFW Freeze'', see Paper I) we aim at constraining the two cosmological parameters, $\Omega_\mathrm{m}$ and $\sigma_8$, plus the two parameters of the $L_x-M$ scaling relation, slope and intercept.
In the discussion section of this paper we estimate the reliability of these masses for the cosmological analysis. 
Our main result for the cosmological analysis incorporates
\begin{itemize}
	\item a lower redshift threshold of $z>0.05$ to exclude galaxy groups on the one hand and the influence of the nearby Universe,
	\item a correction for the extrapolation of the total mass based on the $R_{500}$-test (Section \ref{ch:r500}),
	\item a Gaussian distributed hydrostatic bias of $(1-b)=0.877$ with a standard deviation of $0.015$ as described in \cite{2016ApJ...827..112B} for $\Delta=500$. Since for this overdensity the difference between CC and NCC clusters (as well as for regular and disturbed clusters) is not significant, we use only one bias for all clusters (see also Section \ref{ch:hydrobias}).
	\item The result for $\Omega_\mathrm{m}$ coming from the gas mass fraction analysis is added as a prior in the MCMC.
\end{itemize}
This is our default setup of the mass function analysis. 
Luminosities are taken from \citet{reiprich_hiflugcs} and accounted for the K-correction. 
As pointed out in Paper I, we stay consistent with the selection function of the HIFLUGCS cluster sample by using the same luminosities computed in \cite{reiprich_hiflugcs}, which were mostly computed from ROSAT pointed observations. Moreover, other instruments (Chandra/ACIS, XMM-Newton/EPIC) do not cover the region up to the cluster outskirts. A combined ROSAT and XMM-Newton luminosity comparison has been shown in \cite{zhang_hiflugcs:_2010}, and the authors conclude that the difference to ROSAT-only luminosities is smaller than the intrinsic scatter. Moreover, assuming all our luminosities were offset by a constant factor (as it is found, e.g., by \citealp{zhang_hiflugcs:_2010} for a comparison of ROSAT and XMM-Newton combined luminosities) would not affect our estimates of the cosmological parameters, since a change in the normalization $L_x-M$ relation will compensate for a bias in $L_x$. For a more extended discussion of the $L_x-M$ relation we refer to Paper I.

\begin{figure*}
	\centering
	\resizebox{1.\hsize}{!}{\includegraphics[width=0.95\textwidth]{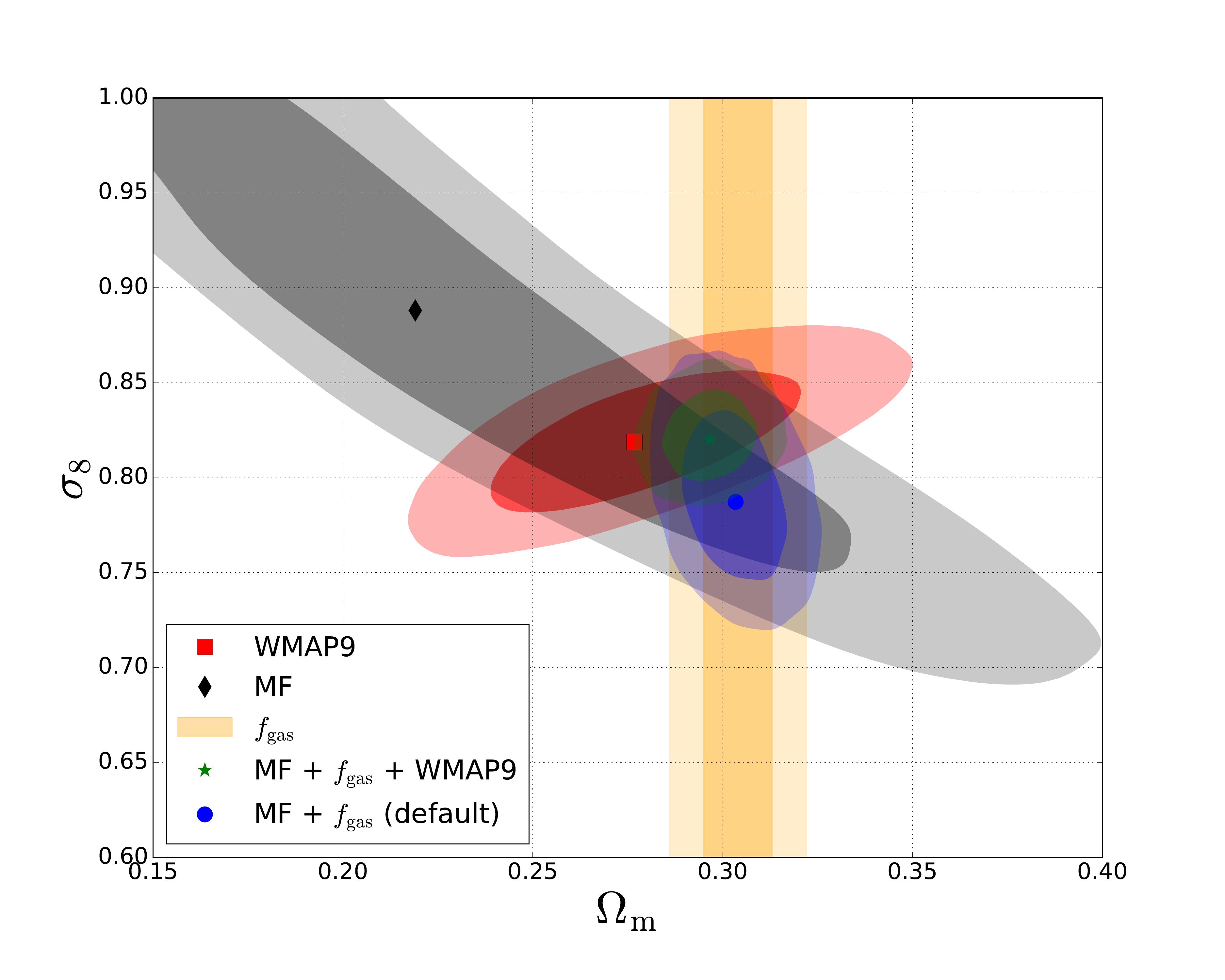}}
	\caption{Results on the cosmological parameters, $\Omega_\mathrm{m}$ and $\sigma_8$. The default setup is shown in blue (circle) which incorporates the \textit{NFW Freeze} masses with the $R_{500}$-test corrections, a hydrostatic bias from \citet{2016ApJ...827..112B}, the lower redshift cut at $\num{0.05}$ and the prior on $\Omega_\mathrm{m}$ from the $f_\mathrm{gas}$ test (yellow). Also shown are the results without the $f_\mathrm{gas}$ constraints (black), the default plus WMAP9 priors (green, WMAP9 alone is shown in red).}
	\label{fig:ellipse_def}
\end{figure*}

\begin{figure*}
	\centering
	\resizebox{1.\hsize}{!}{\includegraphics[width=0.95\textwidth]{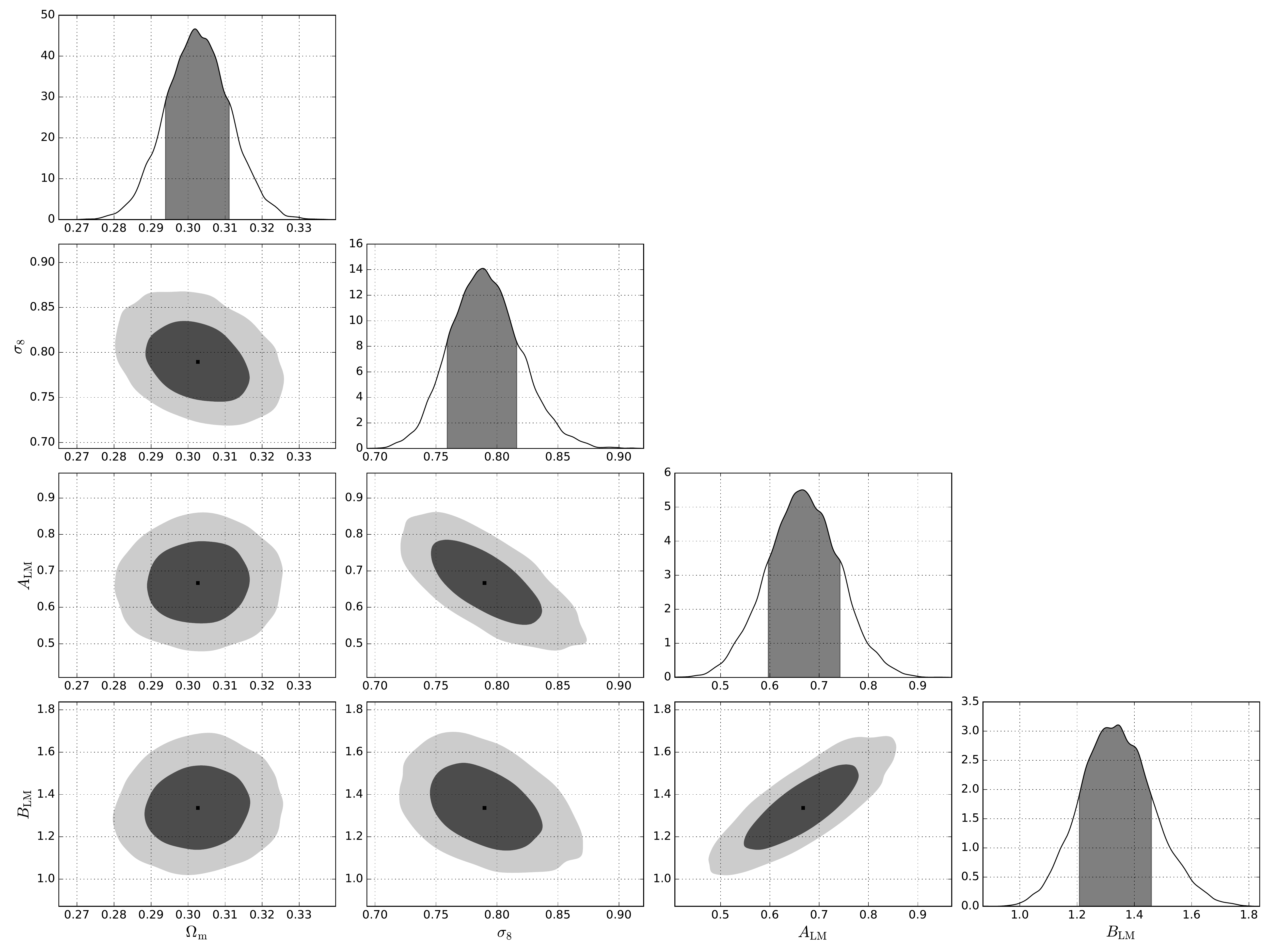}}
	\caption{Confidence levels (68.3\% and 95.4\%) of the 2D parameter space of the four important quantities, $\Omega_\mathrm{m}$, $\sigma_8$, intercept and slope of the $L_x-M$ relation ($A_\mathrm{LM}$ and $B_\mathrm{LM}$, respectively).}
	\label{fig:cosmo_parameters}
\end{figure*}

The scatter of the $L_x-M$ relation is frozen to the observed value, 0.24. The results for the four free  parameters, $\Omega_\mathrm{m}$, $\sigma_8$, $A_\mathrm{LM}$, $B_\mathrm{LM}$, are shown in Fig. \ref{fig:ellipse_def} and Tab. \ref{tab:cosmo_res}.
In Figure \ref{fig:cosmo_parameters} we show the detailed confidence levels (68.3\% and 95.4\%) in the 2D parameter space for the parameter combinations.

\begin{table*}
	\renewcommand{\arraystretch}{1.5}
	\centering
	\begin{tabular}{ccccc}
		\hline
		Setup & $\Omega_\mathrm{m}$ & $\sigma_8$ & $A_\mathrm{LM}$ & $B_\mathrm{LM}$ \\
		\hline
		\hline
		MF + $f_\mathrm{gas}$ (default) &$0.303^{+0.009}_{-0.009}$ & $0.790^{+0.030}_{-0.028}$ & $0.667^{+0.074}_{-0.073}$ & $1.337^{+0.130}_{-0.122}$ \\
		MF + $f_\mathrm{gas}$ + WMAP9 & $0.297^{+0.008}_{-0.008}$ & $0.822^{+0.016}_{-0.014}$ & $0.614^{+0.059}_{-0.058}$ & $1.271^{+0.121}_{-0.110}$ \\
		MF & $0.217^{+0.073}_{-0.054}$ & $0.894^{+0.098}_{-0.095}$ & $0.654^{+0.071}_{-0.071}$ & $1.225^{+0.154}_{-0.154}$ \\
		$f_\mathrm{gas}$ & $0.305^{+0.009}_{-0.009}$ & & & \\
		WMAP9 & $0.279^{+0.027}_{-0.025}$ & $0.821^{+0.024}_{-0.024}$ & & \\
		\hline
		\hline
	\end{tabular}
	\caption{MCMC results for the free parameters from the cosmological pipeline for the different setups. Uncertainties are marginalized 68.3\%.}
	\label{tab:cosmo_res}
\end{table*}

\begin{figure}
	\centering
	\resizebox{1.\hsize}{!}{\includegraphics[width=0.95\textwidth]{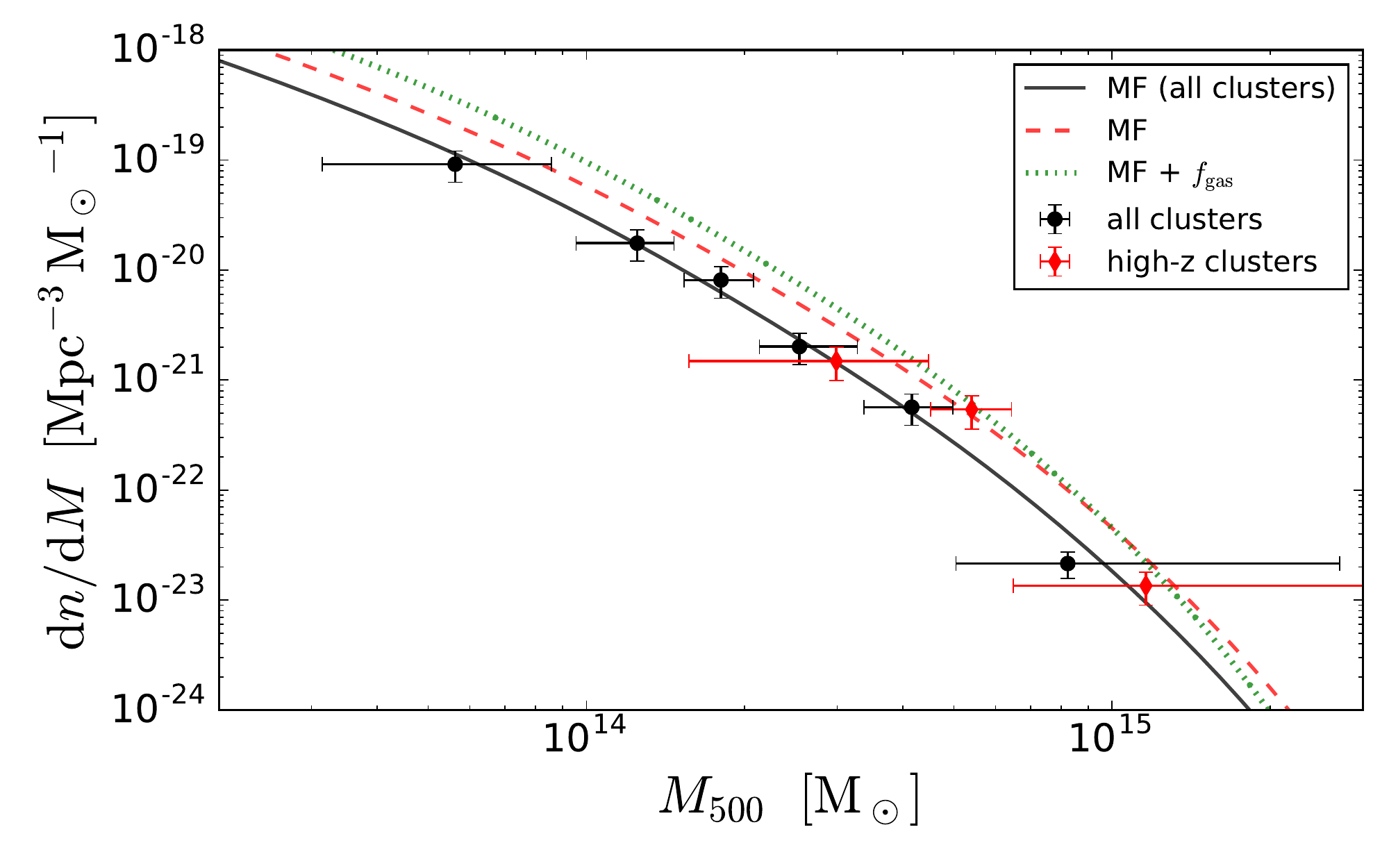}}
	\caption{The cluster mass function (only for illustrational purpose) for the full sample (black) and just the high redshift subsample (red). See text for details.}
	\label{fig:mf_plot}
\end{figure}

The strong degeneracy between $\Omega_\mathrm{m}$ and $\sigma_8$, that can be seen for the mass function (MF) results in Fig. \ref{fig:ellipse_def} disappears when adding prior from the $f_\mathrm{gas}$ test.
For the MF results, the definition of $\sigma_8$ forces the degeneracy, since the root mean square (RMS) amplitude of fluctuations at a given mass, $\sigma(M)$, enters in the mass function and depends on $\Omega_\mathrm{m}$. The limit to a certain scale, $\sigma_8$, still depends on $\Omega_\mathrm{m}$ (see also \citealp{1993MNRAS.262.1023W,2009ApJ...692.1060V}).
\begin{figure}
	\centering
	\resizebox{1.\hsize}{!}{\includegraphics[width=0.49\textwidth]{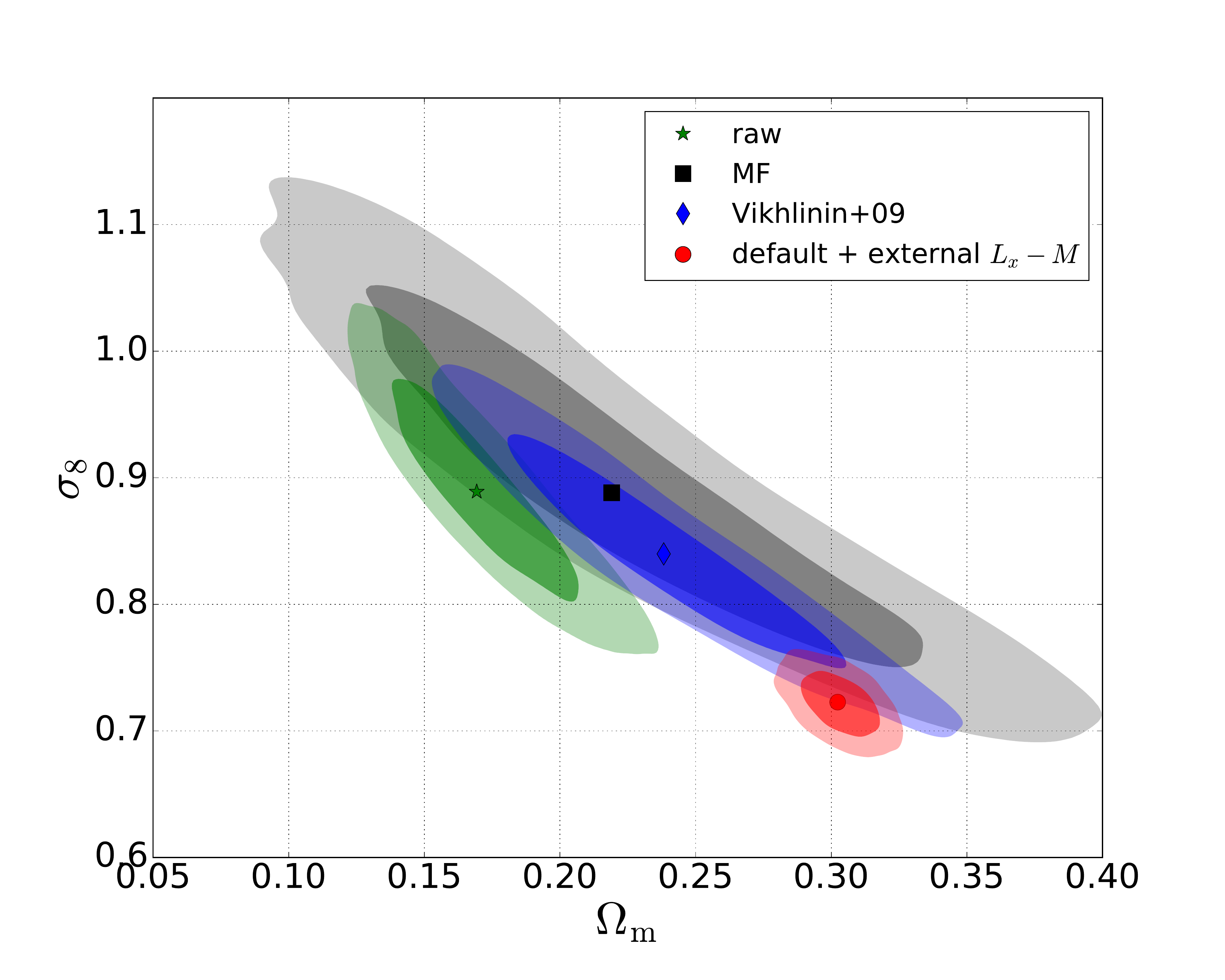}}
	\caption{68.3\% and 95.4\% confidence regions for $\Omega_\mathrm{M}$ and $\sigma_8$ for various analysis setups (see text for details). Note that Vikhlinin$+$09 results have been reanalyzed with our pipeline.}
	\label{fig:oms8_1}
\end{figure}

A correlation between the slope and normalization of the $L_x-M$ relation (the larger the slope, the higher the normalization) is also detected (Pearson 0.81), and does not disappear with a prior on $\Omega_\mathrm{m}$. 
No degeneracy is found for $\Omega_\mathrm{m}$ and the slope or normalization. The chains show a very stable behavior, implying that convergence is very rapid. Testing the chains with different initial values (including values far away from the potential best-fit) produces the same results, which indicates that we are not just mapping a local minimum of the likelihood function.

Figure \ref{fig:mf_plot} shows the halo mass function with the binned cluster masses. The black line and data points correspond to all 64 HIFLUGCS clusters (without redshift cut), the red line and datapoints reflect just the high redshift sample ($z>0.05$) and the green line adds the prior on $\Omega_\mathrm{m}$. 
The masses have been corrected according to the $R_{500}$-test (Section \ref{ch:r500}) and hydrostatic bias, $(1-b) = 0.877$ (Section \ref{ch:hydrobias}).
Note that this illustration does not show a real fit to the datapoints, since the likelihood analysis accounts for each cluster redshift, simultaneously fits the $L_x-M$ relation and does not depend on any binning of clusters.

We tested our pipeline with the (low redshift sub-)sample and results from \citet{2009ApJ...692.1033V,2009ApJ...692.1060V}. This low redshift sample has broad overlap with the HIFLUGCS sample and was constructed using the BCS, REFLEX and HIFLUGCS samples with the same criterion on the survey area as in HIFLUGCS. Fluxes have been redetermined in the $\SIrange{0.5}{2}{keV}$ band using pointed ROSAT observations. The final fluxlimit is $\SI{1.3e-11}{erg\,s^{-1}\,cm^{-2}}$ in this band and additionally a lower redshift limit of 0.025 was applied. This results in 49 galaxy clusters in this low redshift subsample. The high-redshift sample comprises 36 clusters from the 400d survey (\citealp{2007ApJS..172..561B}) above redshift 0.35. In the following we only test the low redshift sample. Masses have been obtained by using either the gas mass $M_\mathrm{gas}$, temperature $kT$, or $Y_x = kT \times M_\mathrm{gas}$ as a proxy. Scaling relations between these quantities and the hydrostatic mass have been calibrated using a low redshift, relaxed sample of 10 clusters (\citealp{2006ApJ...640..691V}). While the actual results on $\Omega_\mathrm{m}$ or $\sigma_8$ do not depend on the choice of the mass proxy, $Y_x$ is chosen as default. The cluster masses $M_{Y_X}$ and luminosities of the low redshift subsample have been taken from this reference and tested with the cosmological analysis pipeline, accounting for the new selection and a new K-correction due to the changed energy band. As seen in Fig. \ref{fig:oms8_1} and Fig. 3 in \citet{2009ApJ...692.1060V}, the results are in good agreement. Note that only the combination of both samples is shown in the reference figure, so perfect agreement is not expected, despite the differences in the analysis strategy (e.g., we assume a flat Universe). 

Another option is to leave the scatter $\sigma_\mathrm{LM}$ free to vary (flat priors). This introduces an additional degree of freedom, but does not change results. The uncertainties increase slightly in this case, but the best-fit scatter is in perfect agreement with the observed scatter, which was used in the default setup.

We perform one more test by fixing the $L_x-M$ relation to the bias-corrected one from \citet{2009A&A...498..361P}, where luminosities were also calculated in the same energy band. In the $\Omega_\mathrm{m} - \sigma_8$ plane the confidence regions follow the main degeneracy of the default case (Fig. \ref{fig:ellipse_def}), but shifted toward lower $\sigma_8$.
We also show in Fig. \ref{fig:oms8_1} the mass function results for the plain \textit{NFW Freeze} masses without $f_\mathrm{gas}$ prior, redshift cut, extrapolation correction and hydrostatic bias (\textit{raw}). Since these results are the most direct, we use these as a baseline in the discussion in the following section. We label those results \textit{raw} from here on, to differentiate from the \textit{default} results, which already include several corrections for systematics.

\begin{figure}
	\centering
	\resizebox{1.\hsize}{!}{\includegraphics[width=0.99\textwidth]{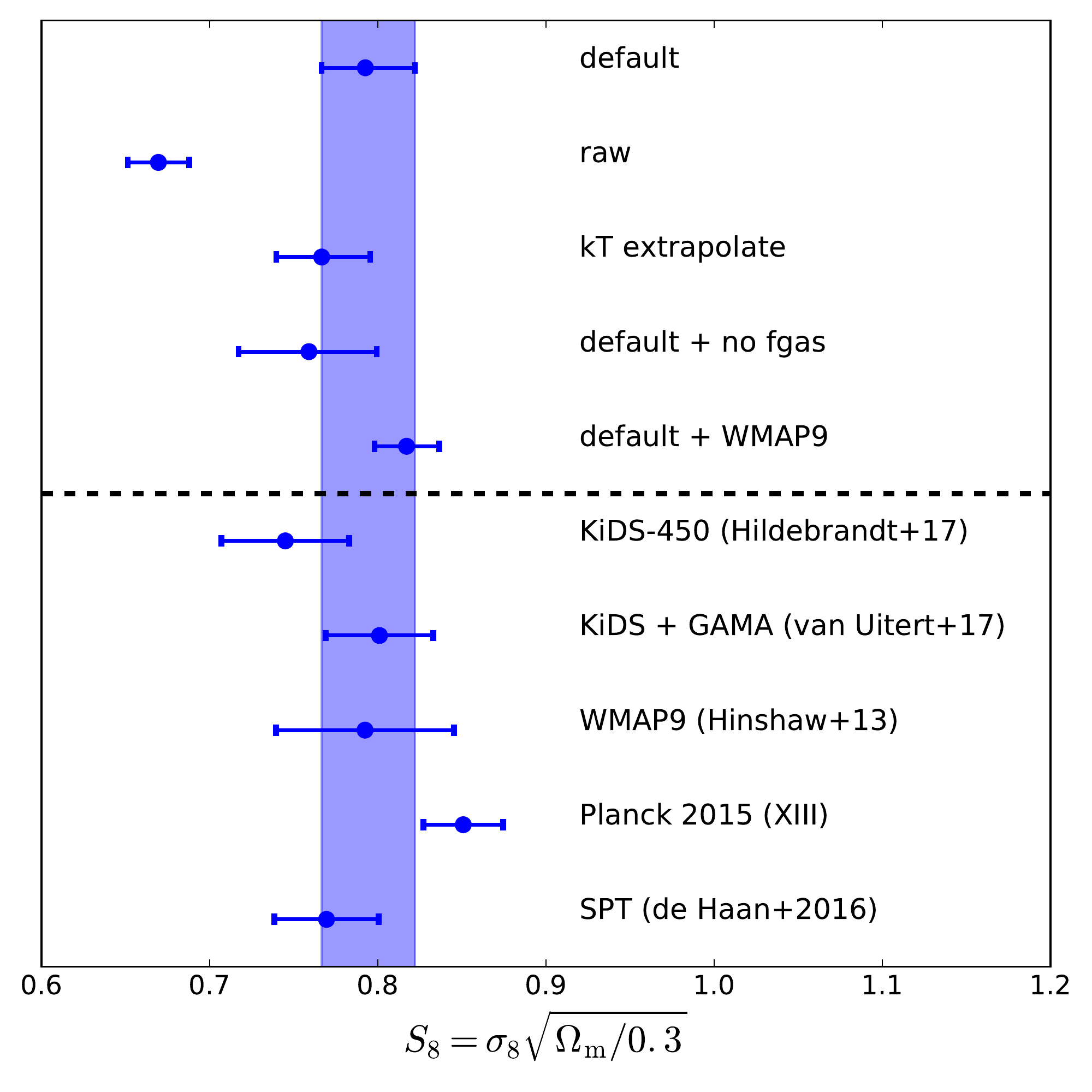}}
	\caption{$S_8$ parameter for various setups and comparison to other cosmological experiments (KiDS: \citealp{2017MNRAS.465.1454H,2017arXiv170605004V}, WMAP9: \citealp{Hinshaw2013}, Planck: \citealp{2015arXiv150201589P}, South Pole Telescope: \citealp{2016ApJ...832...95D}).}
	\label{fig:S8}
\end{figure}
A quantity combining $\Omega_{\rm m}$ and $\sigma_8$, and easily comparable to other experiments is $S_8 = \sigma_8 \sqrt{\Omega_{\rm m} / 0.3}$, which we show in Fig. \ref{fig:S8} for various setups. $S_8$ is sensitive to the location of cosmological constraints in the $\Omega_{\rm m}-\sigma_8$ plane, taking into account the inferred degeneracy of these quantities for mass function analyses. We find excellent agreement of our default result with the WMAP9 (primary anisotropies), SPT (SZ) and KiDS (lensing, especially after combining the with other probes from galaxy-galaxy-lensing and angular clustering, see \citealp{2017arXiv170605004V}) results. Our raw results without any corrections are not in agreement. The source of this bias is evaluated in the following section.
\section{Discussion}
\label{ch:hicosmo_discussion}
In this section we discuss the systematic effects that might enter in our cosmological analysis. 
We compare here the effect of every test with the ``raw'' results, rather than the ``default'' corrected results. 
In Table \ref{tab:cosmo_res2} we summarize the cosmological results of the various tests and also estimate their bias by computing the shift in $S_8$ with respect to the ``raw'' case.

\subsection{Difference to WMAP9}
\label{ch:diff_wmap9}
\begin{figure}
	\centering
	\resizebox{1.\hsize}{!}{\includegraphics[width=0.49\textwidth]{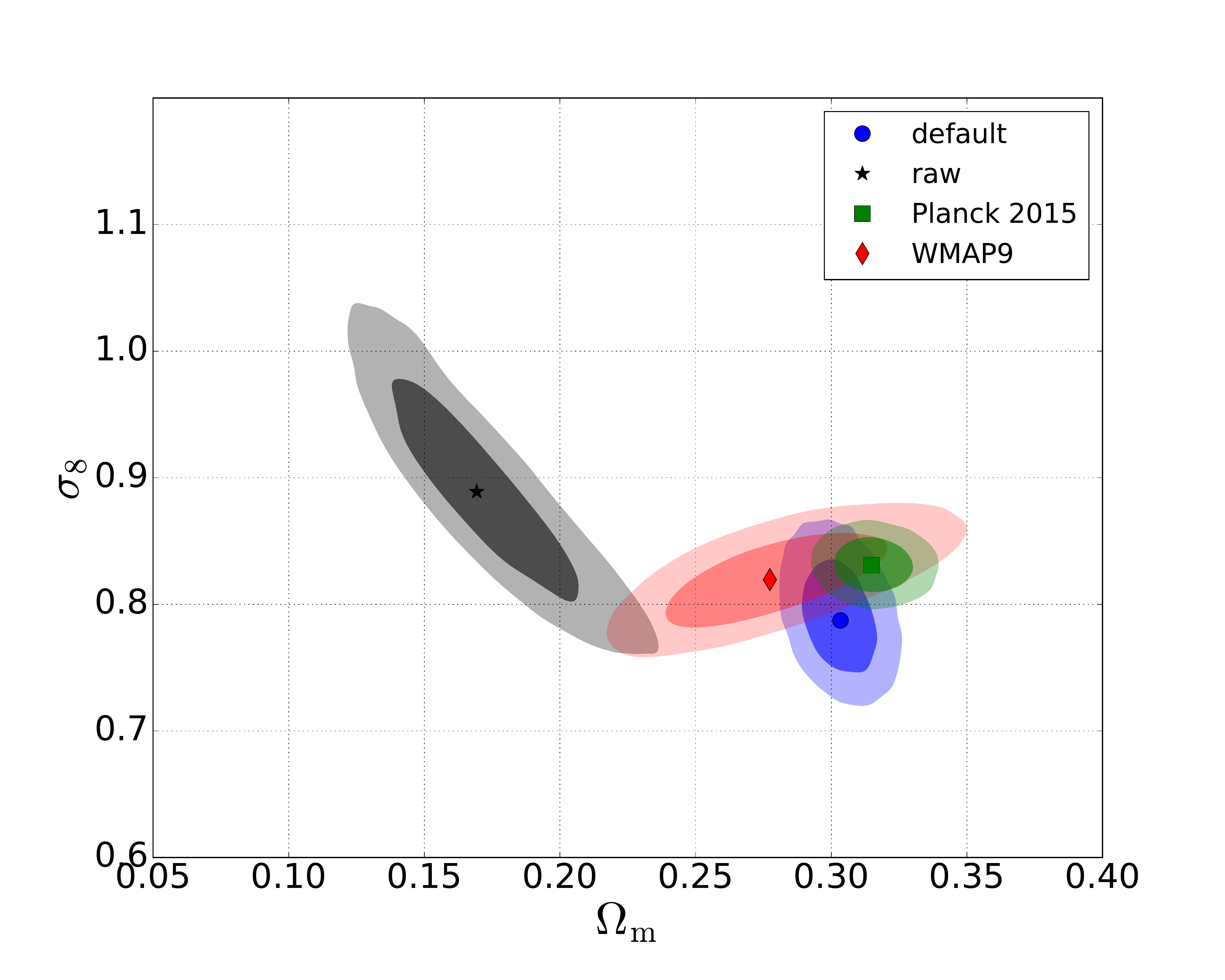}}
	\caption{HIFLUGCS cosmology results compared to WMAP9 (\citealp{Hinshaw2013}) and Planck 2015 (\citealp{2015arXiv150201589P}) results in the $\Omega_\mathrm{m}$-$\sigma_8$ plane. The contours correspond to 68.3\% and 95.4\% confidence regions.}
	\label{fig:oms8_2}
\end{figure}

CMB experiments which measure temperature fluctuations at a large variety of angular scales can constrain several cosmological parameters with great precision and provide, independent of galaxy clusters, another reference for the composition and evolution of the Universe.
The two latest all-sky CMB temperature fluctuation measurements come from the WMAP satellite (9 year data; \citealp{Hinshaw2013}) and the Planck Satellite (\citealp{2015arXiv150201589P}) and are compared to the HIFLUGCS cosmological results in Fig. \ref{fig:oms8_2}. 
One great advantage of combining different measurements is to eliminate the degeneracy, e.g., of $\Omega_\mathrm{m}$ and $\sigma_8$. 
The WMAP9 and HIFLUGCS (raw) results exhibit small overlap, while the Planck 2015 results are shifted toward larger $\Omega_\mathrm{m}$. In the following we focus on a comparison with the WMAP9 results, for the following reasons: A Planck cosmology would predict slightly too many galaxy clusters (e.g., \citealp{2015arXiv151204264P}) and there is some tension with the Planck SZ results, which predict smaller $\Omega_\mathrm{m}$. The latter problem could be solved at least partially by using a weak lensing mass calibration (\citealp[Fig. 7]{PlanckCollaboration2015ae}). Furthermore, the Planck CMB result itself is not consistent between the low and high multipole constraints (\citealp{2015arXiv151100055A}). 

The results of the combined MF+$f_\mathrm{gas}$+WMAP9 analysis, which is the default setup with the WMAP9 constraints as priors on $\Omega_\mathrm{m}$ and $\sigma_8$, are also shown in Tab. \ref{tab:cosmo_res} and Fig. \ref{fig:ellipse_def}.
\begin{table*}
	\renewcommand{\arraystretch}{1.5}
	\centering
	\begin{tabular}{ccccccc}
		\hline
		Setup & $\Omega_\mathrm{m}$ & $\sigma_8$ & $A_\mathrm{LM}$ & $B_\mathrm{LM}$ & $\sigma_\mathrm{LM}$ & $\Delta S_8$ \\
		\hline
		\hline
		default & \multirow{2}{*}{$0.303^{+0.009}_{-0.009}$} & \multirow{2}{*}{$0.790^{+0.030}_{-0.028}$} & \multirow{2}{*}{$0.667^{+0.074}_{-0.073}$} & \multirow{2}{*}{$1.337^{+0.130}_{-0.122}$} & \multirow{2}{*}{$0.24$} & \multirow{2}{*}{$+0.123$} \\
		
		[raw + $R_{500}$-test + $(1-b)=$Biffi16 + $z>0.05$] &  &  &  &  & & \\		
		\hline
		default + WMAP9 & $0.297^{+0.008}_{-0.008}$ & $0.822^{+0.016}_{-0.014}$ & $0.614^{+0.059}_{-0.058}$ & $1.271^{+0.121}_{-0.110}$ & $0.24$ & $+0.148$\\
		\hline
		NFW Freeze (raw) & $0.170^{+0.024}_{-0.021}$ & $0.888^{+0.053}_{-0.052}$ & $0.842^{+0.062}_{-0.059}$ & $1.351^{+0.073}_{-0.071}$  & $0.26$ & -- \\
		\hline
		NFW All (raw-like) & $0.216^{+0.024}_{-0.023}$ & $1.085^{+0.036}_{-0.038}$ & $0.136^{+0.056}_{-0.054}$ & $1.204^{+0.048}_{-0.048}$  & $0.42$ & $+0.249$ \\	
		\hline
		NFW Hudson (raw-like) & $0.203^{+0.025}_{-0.020}$ & $1.087^{+0.038}_{-0.045}$ & $0.196^{+0.055}_{-0.054}$ & $1.213^{+0.053}_{-0.051}$ &$0.42$ & $+0.223$ \\	
		\hline
		kT extrapolate (raw-like) &  $0.175^{+0.027}_{-0.023}$ & $1.003^{+0.071}_{-0.068}$ & $0.479^{+0.067}_{-0.066}$ & $1.234^{+0.078}_{-0.076}$ & $0.30$ & $+0.097$ \\
		\hline
		XMM-Newton (kT extr.) & $0.167^{+0.022}_{-0.019}$   &  $0.853^{+0.048}_{-0.057}$  & $0.904^{+0.069}_{-0.061}$   & $1.316^{+0.077}_{-0.065}$   & $0.24$ & $-0.039$  \\
		\hline
		raw + $z>0.05$ & $0.200^{+0.065}_{-0.047}$ & $0.889^{+0.088}_{-0.091}$ & $0.696^{+0.071}_{-0.068}$ & $1.182^{+0.142}_{-0.131}$ & $0.24$ & $+0.056$ \\
		\hline
		raw + $z<0.05$ & $0.150^{+0.039}_{-0.028}$ & $0.914^{+0.134}_{-0.121}$ & $1.244^{+0.107}_{-0.106}$ & $1.627^{+0.101}_{-0.100}$ & $0.22$ & $-0.025$ \\
		\hline
		raw + 25\% Groups & $0.223^{+0.033}_{-0.029}$ & $0.820^{+0.057}_{-0.052}$ & $0.767^{+0.057}_{-0.056}$ & $1.242^{+0.072}_{-0.069}$ & $0.26$ & $+0.039$ \\
		\hline
		raw + 50\% Groups & $0.198^{+0.028}_{-0.026}$ & $0.854^{+0.058}_{-0.053}$ & $0.796^{+0.059}_{-0.056}$ & $1.291^{+0.072}_{-0.070}$ & $0.26$ & $+0.024$ \\
		\hline
		raw + 80\% Groups & $0.178^{+0.024}_{-0.022}$ & $0.884^{+0.054}_{-0.052}$ & $0.811^{+0.058}_{-0.058}$ & $1.318^{+0.069}_{-0.066}$ & $0.26$ & $+0.012$ \\
		\hline
		raw + Broken Powerlaw & $0.141^{+0.021}_{-0.019}$ & $0.988^{+0.072}_{-0.068}$ & $0.661^{+0.133}_{-0.132}$ & $1.697^{+0.128}_{-0.122}$ & ${}^\mathrm{a}0.995^{+0.112}_{-0.106}$ & $+0.009$ \\
		\hline
		raw + $(1-b) = [0.7,1]$ & $0.174^{+0.026}_{-0.021}$ & $0.938^{+0.066}_{-0.061}$ & $0.725^{+0.082}_{-0.079}$ & $1.336^{+0.071}_{-0.068}$ & $0.26$ & $+0.046$\\
		\hline
		raw + $(1-b) = 0.8$ & $0.177^{+0.025}_{-0.021}$ & $0.950^{+0.058}_{-0.055}$ & $0.694^{+0.054}_{-0.052}$ & $1.338^{+0.070}_{-0.068}$ & $0.26$ & $+0.059$ \\
		\hline
		raw + $z>0.05$ +  & \multirow{2}{*}{$0.280^{+0.023}_{-0.020}$} & \multirow{2}{*}{$0.822^{+0.021}_{-0.019}$} & \multirow{2}{*}{$0.635^{+0.089}_{-0.090}$} & \multirow{2}{*}{$1.281^{+0.123}_{-0.111}$} & \multirow{2}{*}{$0.26$} & \multirow{2}{*}{$+0.124$}\\
		
		$(1-b)=[0.7,1]$ + WMAP9 & & & & & & \\
		\hline
		raw + undisturbed & $0.213^{+0.034}_{-0.028}$ & $0.784^{+0.056}_{-0.052}$ & $0.981^{+0.073}_{-0.075}$ & $1.516^{+0.086}_{-0.086}$ & $0.26$ & $-0.012$ \\
		\hline
		raw + undisturbed + $(1-b) = 0.8$ & $0.224^{+0.036}_{-0.031}$ & $0.828^{+0.057}_{-0.057}$ & $0.837^{+0.064}_{-0.066}$ & $1.524^{+0.085}_{-0.087}$ & $0.26$ & $+0.045$ \\
		\hline
		raw + undisturbed +  & \multirow{2}{*}{$0.271^{+0.023}_{-0.021}$} & \multirow{2}{*}{$0.816^{+0.021}_{-0.021}$} & \multirow{2}{*}{$0.727^{+0.119}_{-0.104}$} & \multirow{2}{*}{$1.581^{+0.064}_{-0.061}$} & \multirow{2}{*}{$0.26$} &\multirow{2}{*}{$+0.106$}\\
		
		$(1-b)=[0.7,1]$ + WMAP9 & & & & & & \\
		\hline
		Planck SZ Masses & $0.234^{+0.042}_{-0.033}$ & $0.790^{+0.051}_{-0.052}$ & $0.980^{+0.055}_{-0.053}$ & $1.606^{+0.107}_{-0.095}$ & $0.197^{+0.025}_{-0.021}$ & $+0.028$ \\
		\hline
		Dynamical Masses & $0.171^{+0.027}_{-0.021}$ & $0.944^{+0.063}_{-0.062}$ & $0.573^{+0.068}_{-0.072}$ & $1.236^{+0.079}_{-0.078}$ & $0.35$ & $+0.050$ \\
		\hline
		raw + Bocquet DM & $0.163^{+0.024}_{-0.021}$ & $0.857^{+0.041}_{-0.041}$ & $0.827^{+0.056}_{-0.055}$ & $1.339^{+0.069}_{-0.070}$ & $0.26$ & $-0.038$ \\
		\hline
		raw + Bocquet Hydro & $0.171^{+0.024}_{-0.021}$ & $0.845^{+0.038}_{-0.039}$ & $0.822^{+0.057}_{-0.057}$ & $1.334^{+0.066}_{-0.065}$ & $0.26$ & $-0.031$ \\
		\hline
		raw + $\sum m_\nu = 0.5\,\mathrm{eV}$ & $0.187^{+0.024}_{-0.020}$ & $0.850^{+0.045}_{-0.045}$ & $0.827^{+0.060}_{-0.058}$ & $1.343^{+0.066}_{-0.071}$ & $0.26$ & $+0.002$\\
		\hline
		raw + $\sum m_\nu = 1.0\,\mathrm{eV}$ & $0.215^{+0.026}_{-0.023}$ & $0.794^{+0.040}_{-0.039}$ & $0.839^{+0.059}_{-0.059}$ & $1.354^{+0.070}_{-0.069}$ & $0.26$ & $+0.003$\\
		\hline
		\hline
	\end{tabular}
	\caption{MCMC results for the free parameters from the cosmological pipeline for the default setup and various setups used to test systematic effects. Uncertainties are purely statistical and marginalized at 68.3\%. $x$\% Groups refer to the normal skyfraction multiplied by $x$/100 for all objects with $M < \SI{e14}{M_\odot}$. a: The high mass slope for the broken powerlaw instead of the scatter, which is $0.26$. \textit{raw-like} means that the setup is identical to the \textit{raw} case, only the extrapolation method is different. The last column shows the difference in the derived parameter $S_8$ (see Fig. \ref{fig:S8}) with respect to the ``raw'' setup.}
	\label{tab:cosmo_res2}
\end{table*}
The uncertainties of $\sigma_8$ decrease by about 50\%, while the uncertainties of $\Omega_\mathrm{m}$ are already limited by the $f_\mathrm{gas}$ prior. 

\subsection{Systematics of the mass determination}
It is of crucial importance to take all possible effects into account, which could systematically bias the results. In the following we will discuss the influence of the extrapolation methods, instrumental calibration effects and the use of a different overdensity than the default $500\rho_\mathrm{crit}$. This will lead to an empiric correction for the extrapolation based on clusters where the temperature profile is measured out to $R_{500}$.
Physical effects in individual clusters, such as strong AGN feedback, and modifications to the cosmological model, are discussed in subsections (\ref{ch:disc_gcphysics} and \ref{ch:hicosmo_model}).

One general concern connected with the likelihood function and the sample selection is that a few clusters are very close to the flux limit. There is a chance that these clusters might slip, due to the variable K-correction (we compute the fluxes from the given luminosities), slightly below the flux limit, which results in a very small value of the likelihood. By excluding\footnote{In order to not change the statistics these clusters were not really excluded, but just given much larger uncertainties on the luminosity.} these clusters from the analysis it was ensured that the cosmological results do not get biased by a very small number of clusters at the flux limit. 

A similar effect could also happen if a cluster is far away from the best fit $L_x-M$ relation. Usually the luminosity uncertainties are very small and the probability could essentially be $0$ or very close to it. Excluding these objects as well ensures that distant outliers are not influencing the results too much. As with the lowest flux objects, here we also conclude that $L_x-M$ relation outliers are not biasing the results.

\subsubsection{Impact of different extrapolation methods}
\label{ch:hicosmo_disc_extra}
\begin{figure}
	\centering
	\resizebox{1.\hsize}{!}{\includegraphics[width=0.49\textwidth]{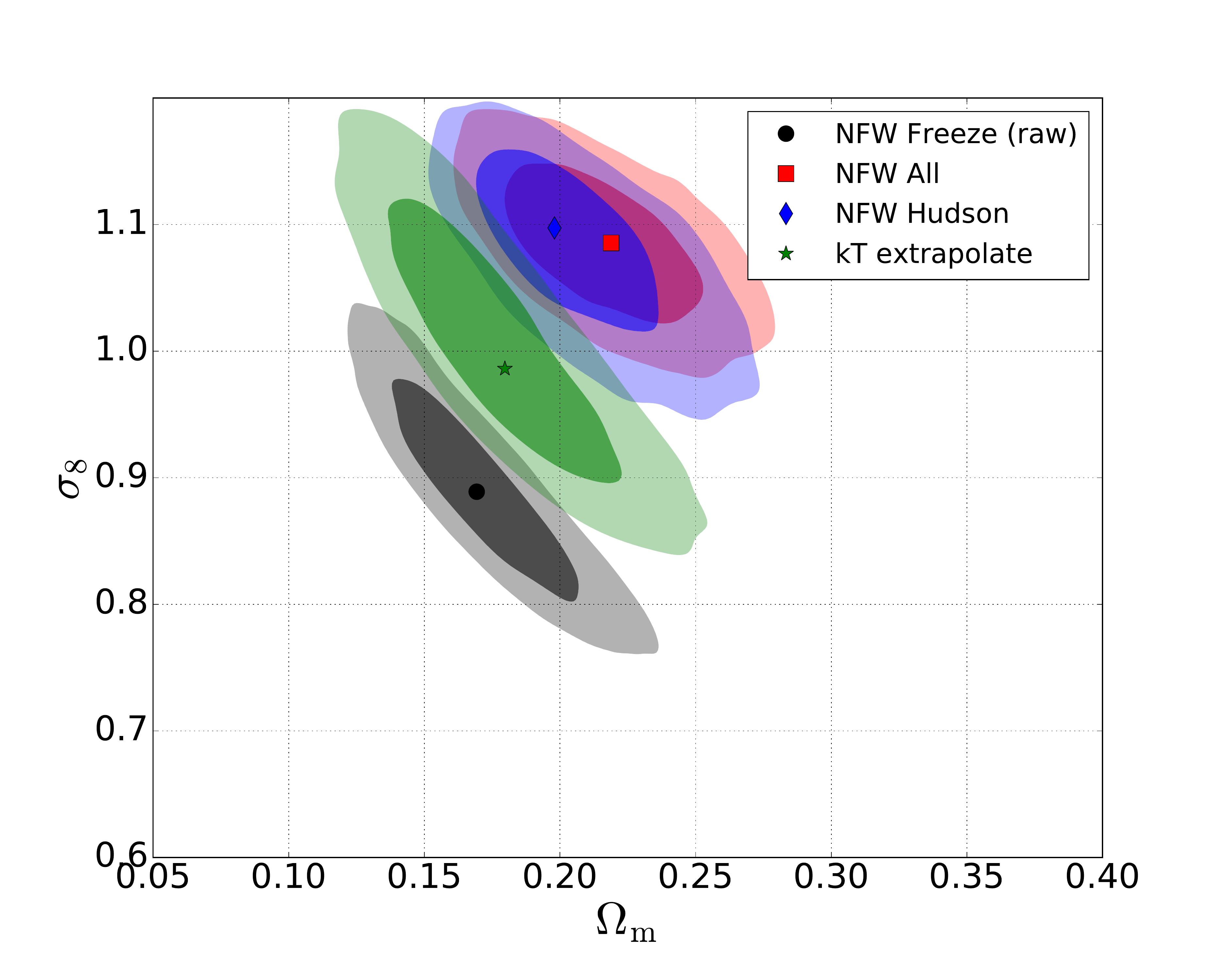}}
	\caption{68.3\% and 95.4\% confidence levels for $\Omega_\mathrm{m}$ and $\sigma_8$ for the different extrapolation methods (see also Tab. \ref{tab:cosmo_res2}).}
	\label{fig:mass_comps}
\end{figure}

In Paper I, four different total mass estimates were introduced, of which the ``NFW Freeze'' method was chosen as default and used up to now. Although the default method should give the most robust mass estimates for large extra\-polations, we show the cosmological results using the other mass estimates as well. Since some of the ``NFW All'' or ``NFW Hudson'' masses are unphysically high ($> \SI{e16}{M_\odot}$), the results might produce biased $\sigma_8$ values, since this parameter is sensitive to the high mass end of the mass function. 
Figure \ref{fig:mass_comps} shows the cosmological constraints of these four different methods to obtain masses.
The temperature profile extrapolation method (green) seems to be almost in agreement with constraints of the standard method (black) and gives very similar results in the $\Omega_\mathrm{m}-\sigma_8$ plane. The uncertainties are slightly larger for ``kT extrapolate'', since ``NFW Freeze'' has only one free parameter for the extrapolation. 

The results of ``NFW All'' and ``NFW Hudson'' are in very good agreement. Both methods use the inner parts of the mass profile (``Hudson'' excludes the central cool core region, if present) for the extrapolation. The inner regions might deviate from an NFW profile, having a steeper central profile slope. This causes some cluster masses to be biased high, since the concentration parameter of the NFW profile becomes unrealistically small ($< 1$) to fit the profile. In particular, an intrinsically high mass at the high mass end will force  larger $\sigma_8$, which can be seen in Fig. \ref{fig:mass_comps}. 
If we ``average'' the four masses for each cluster, by taking the median and using the highest and lowest mass as the upper and lower uncertainty range, we find that the uncertainties of $\Omega_\mathrm{m}$ and $\sigma_8$ increase by roughly 30\%, which reflects the systematics involved in the extrapolation.

\subsubsection{XMM-Newton masses}
In \cite{Schellenberger2015} the impact of instrumental cross calibration uncertainties on the cosmological constraints from the mass function has been evaluated. The temperature
scaling \cite[Table 2]{Schellenberger2015} was used to convert Chandra derived temperature profiles into XMM-Newton
profiles, which is the input for the mass determination and cosmological analysis. The main result was
a shift toward slightly smaller $\Omega_{\rm m}$ for the XMM-Newton masses. While in \cite{Schellenberger2015} galaxy clusters
have been binned in mass and an external $L_x-M$ relation from \cite{reiprich_hiflugcs} was
used to calculate the volumes, here we re-evaluate this effect using the derived masses for HIFLUGCS
and the likelihood approach including the simultaneous fit of the $L_x-M$ relation. For the scaling of
the temperature profiles we again use the ``ACIS-Combined XMM'' relation for the full energy band. We note out that the Chandra calibration has changed, and now includes the new contamination model
vN0008, but the effect of this change is small (< 2\%). 
For this test we focus on the temperature
extrapolation method used to calculate the masses, because the purpose is to study the effects of
calibration uncertainties (entering in the temperature profile) and NFW fits could possibly introduce an
additional bias.
We find Chandra and scaled XMM-Newton masses exhibit a different
behavior than what was inferred in the previous work: In \cite{Schellenberger2015} the masses were
found to have a constant fractional difference ($\sim 14\%$), while now the difference between Chandra and
XMM-Newton masses is increasing with mass, which leads to lower $\sigma_8$ of the XMM-Newton masses with respect to Chandra. 
We confirm the previous results that the overall shift in the
$\Omega_{\rm m}-\sigma_8$ plane cannot explain the difference between cosmological constraints of Planck primary CMB
anisotropies and SZ, and also that for the present study, the shift is smaller than the statistical uncertainty.

\subsubsection{Different overdensities}
The ideal halo mass function is a universal parametrization, which applies at all redshifts and cosmologies. Unfortunately, as pointed out in \citet{2008ApJ...688..709T,Bocquet2015a}, not only redshift correction have to be made, but also at overdensities larger than $180 \rho_\mathrm{mean}$, deviations from universality should be expected. Many cosmological analyses that involve galaxy clusters use $500 \rho_\mathrm{crit}$ as an overdensity for the mass calculation, which enables us to easily compare our results. Furthermore, $500\rho_\mathrm{crit}$ seems to be a good compromise for keeping extrapolation to a minimum and not moving to too small radii, where hydrostatic equilibrium might not hold.
We recalculate the masses and extrapolate until $200\rho_\mathrm{crit}$ using the ``kT extrapolate'' and ``NFW Freeze'' methods. 
For the temperature extrapolation no significant change can be detected, only that the uncertainties increase. For the ``NFW Freeze'' method, the uncertainties increase as well, but there is also a clear shift toward higher ($\sim 40\%$) $\Omega_\mathrm{m}$. This can be explained because the NFW model with a frozen concentration parameter puts more constraints on the shape of the mass profile, which can cause larger biases the more it is extrapolated. The observed direction of the shift toward larger $\Omega_\mathrm{m}$ may just be chance.

\subsubsection{$R_{500}$-test}
\label{ch:r500}
Here we focus on the systematics of the default extrapolation method, \textit{NFW Freeze}. Starting from 6 clusters\footnote{A85, A1644, A2029, A3667, A4059 and HydraA} where the temperature profile can be measured out to more than 90\% of $R_{500}$, we recalculate the total mass by removing all temperature measurements beyond 20\%, 40\%, 60\% and 80\% of the real $R_{500}$. 
Despite the scatter among the six clusters, we find that with smaller covered regions of $R_{500}$ also the total mass is underestimated (using the \textit{NFW Freeze} method).
We parameterize the fraction of the real $M_{500}$, $\kappa$, as a function of the measured $R_{500}$ fraction, $\epsilon$:
\begin{equation}
	\kappa = 1 + e^{-a} - e^{-a \epsilon^b}~,
\end{equation}
where $a=9.7$ and $b=1.7$. 
This means if only half of the temperature profile is measured, the mass is underestimated by 5\%, while if only 30\% is measured one will also find a 30\% too low mass.
The scatter (standard deviation, $\Delta \kappa$) between the results of the 6 individual clusters is used as an additional uncertainty and measured to follow the relation
\begin{equation}
	\Delta \kappa = 0.31\cdot(1-\epsilon)~.
\end{equation}
For the default results, all our measured masses are corrected by a factor $1/\kappa$, while the $\Delta \kappa$ is added in quadrature to the relative uncertainty of the cluster mass.

\subsection{Galaxy cluster physics}
\label{ch:disc_gcphysics}
In this Section we will discuss various effects and processes that could cause the observed difference in $\Omega_\mathrm{m}$ and $\sigma_8$, but we do point out that the aim is not to reproduce, e.g., the CMB results, but to test several influences that enter in a purely flux limited sample.
\subsubsection{Galaxy groups}
\label{ch:groups}
\begin{figure*}
	\centering
	\resizebox{1.\hsize}{!}{\includegraphics[width=0.8\textwidth]{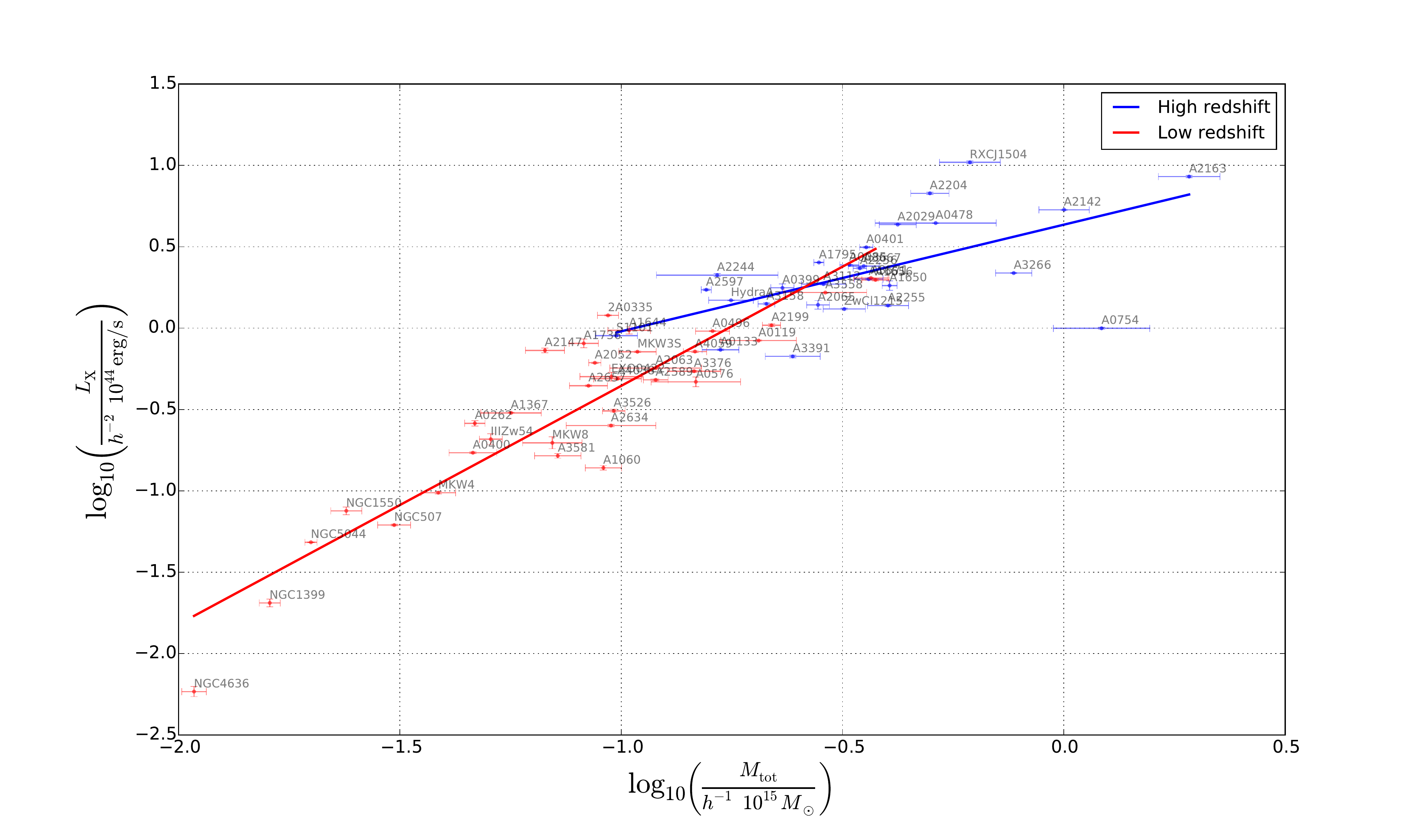}}
	\caption{$L_x-M$ relations for the low (raw + $z < 0.05$) and high redshift (raw + $z \geq 0.05$) sample.}
	\label{fig:gcphysics_1}
\end{figure*}

\begin{figure*}
	\centering
	\resizebox{1.\hsize}{!}{\includegraphics[width=0.8\textwidth]{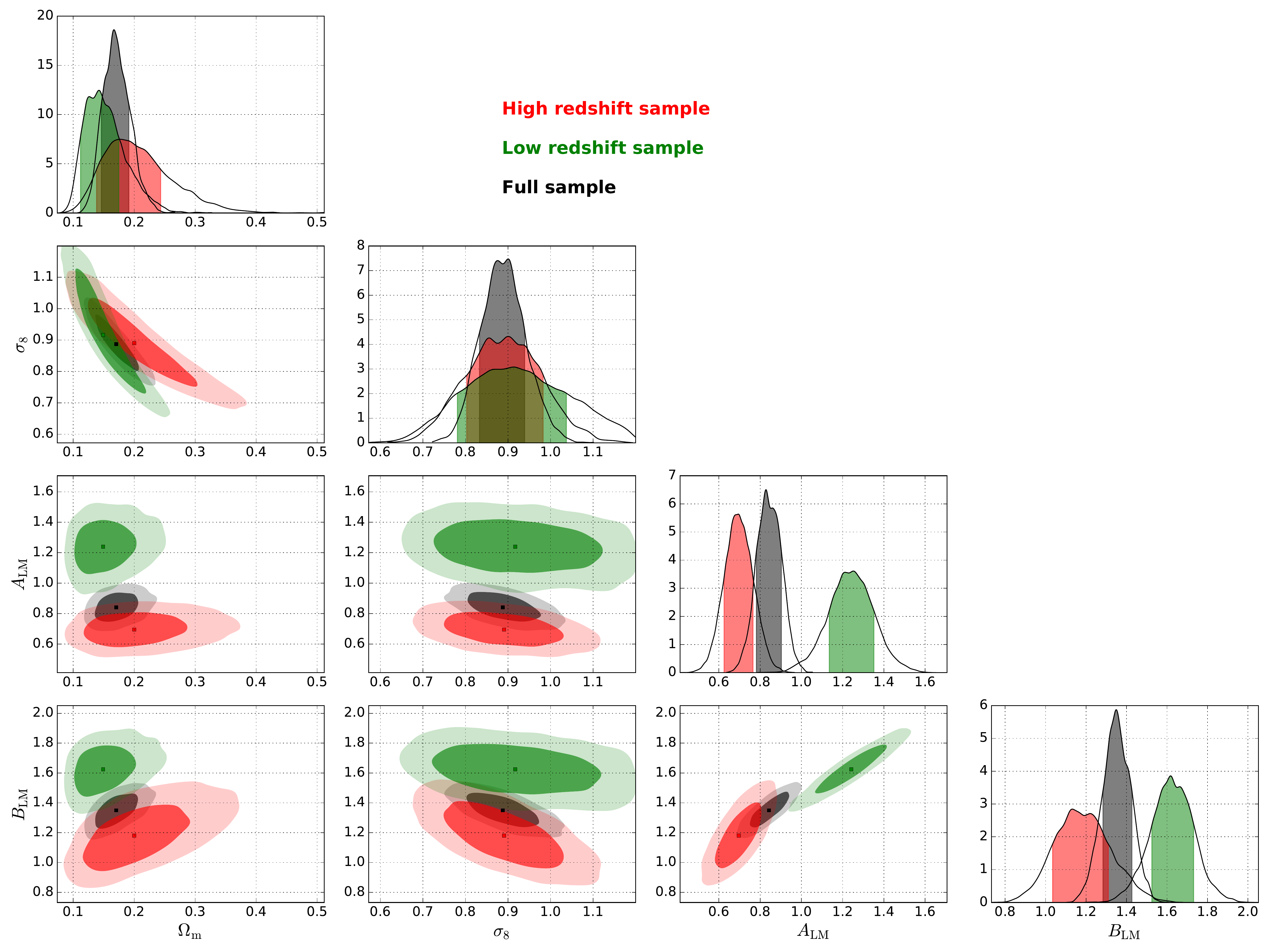}}
	\caption{68.3\% and 95.4\% confidence levels for the full sample (black, raw in Tab. \ref{tab:cosmo_res2}), the high redshift sample (raw + $z \geq 0.05$, red) and the low redshift sample (raw + $z<0.05$, green).}
	\label{fig:gcphysics_2}
\end{figure*}

The first effect to be analyzed here is the sample composition: In contrast to, e.g., \citet{2009ApJ...692.1060V}, the HIFLUGCS sample includes several galaxy groups. 
In the following we consider every object with $M_{500} < \SI{e14}{\textit{h}^{-1}\,M_\odot}$ as a galaxy group, since there is not a well defined threshold to separate groups from clusters (\citealp{2009ApJ...693.1142S}).
As shown, e.g., in \citet{2009ApJ...693.1142S,2011A&A...535A.105E,Bharadwaj2014,Lovisari2015} these objects have different scaling properties than galaxy clusters. One simple powerlaw to describe the $L_x-M$ relation for the full sample might not be enough. First, we exclude the galaxy groups by introducing a lower redshift cut. We decided to use 0.05 as the redshift threshold; this will split the sample into two equal sized subsamples and the minimum mass of the high redshift samples is roughly $\SI{e14}{\textit{h}^{-1}\,M_\odot}$ (see Fig \ref{fig:gcphysics_1}). Already the observed $L_x-M$ relations of these subsamples show a clear break with the high redshift (and high mass) sample found to be significantly flatter.
$\Omega_\mathrm{m}$ and $\sigma_8$ can be seen in Fig. \ref{fig:gcphysics_2}: The high redshift sample gives larger values of $\Omega_\mathrm{m}$, which cannot be explained by the degeneracy between the $L_x-M$ slope and $\Omega_\mathrm{m}$. The $\Omega_\mathrm{m}-\sigma_8$ constraints are in perfect agreement with \citet{2009ApJ...692.1060V}, but have larger uncertainties. 

To test in more detail the effect on the mass function of galaxy groups in the sample we introduced a scaling factor $x$ for the skyfraction (by default set to $64.78\%$) for objects with a mass lower than $\SI{e14}{M_\odot}$ (for $h = 0.71$). This should mimic a possible increase in the incompleteness of ROSAT catalogs on galaxy group scale. Implicitly this also tests if missing flux for galaxy groups can be an issue for the cosmological constraints of the HIFLUGCS sample. For example in \citet{Lovisari2015}, a higher luminosity for low mass systems was detected than what is given in the ROSAT catalogs. The authors argue that ROSAT was not able to detect the emission out to large radii for these faint objects. In order not to have discontinuities we model this change in the skyfraction by a sigmoid function, where 99\% of the final skyfraction is reached at mass of $\SI{1.7e14}{M_\odot}$.
\begin{figure}
	\centering
	\resizebox{1.\hsize}{!}{\includegraphics[width=0.99\textwidth]{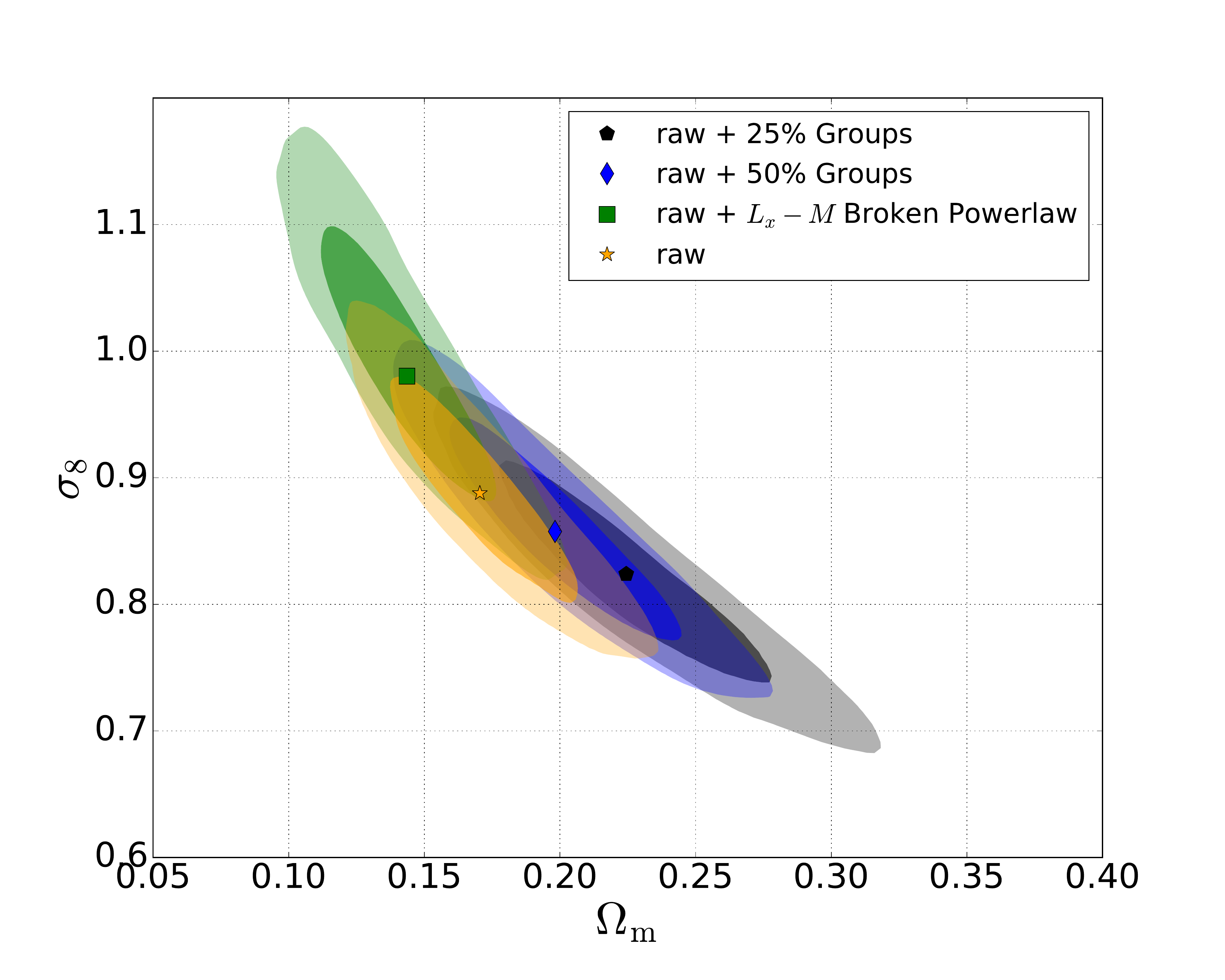}}
	\caption{68.3\% and 95.4\% confidence levels for $\Omega_\mathrm{m}$ and $\sigma_8$ assuming higher incompleteness of galaxy groups in the catalogs (black and blue), or an broken powerlaw for the $L_x-M$ relation (green).}
	\label{fig:groups}
\end{figure}
We set $x$ to 25\%, 50\% and 80\%. For 80\% and 50\% we do not detect any significant change in the cosmological parameters, while for 25\% of the original skyfraction there is a clear shift toward higher $\Omega_\mathrm{m}$ (see Tab. \ref{tab:cosmo_res2} and Fig. \ref{fig:groups}), which seems to be in rough agreement with WMAP9 results. However, detecting only a quarter of the existing galaxy groups is quite unrealistic, given the high flux limit, where even low mass objects are reliably detected (e.g., \citealp[Fig. 23]{2001A&A...369..826B}). 

Another approach to model the different behavior of galaxy groups is to use a broken powerlaw for the $L_x-M$ relation (Tab. \ref{tab:cosmo_res2}). We set the break point for the slope to $\SI{e14}{M_\odot}$. As expected a much shallower slope is detected for the high mass objects, which is only about 50\% of the slope for the low mass objects. The result for $\Omega_\mathrm{m}$ and $\sigma_8$ is only shifted along the degeneracy toward lower $\Omega_\mathrm{m}$. But in case groups are missed in the sample due to selection effects or catalog incompleteness a broken powerlaw would just model the observed behavior and neglect these effects. The real distribution of galaxy clusters and groups in the $L_x-M$ plane could look different, and unknown effects push the groups toward lower luminosities, which cause the observed steepening. So a lower $\Omega_\mathrm{m}$, mostly driven by the lower number of groups that has to be matched, is the expected trend for this case.
The split into a high and low redshift sample seems to be more justified than to just treat groups differently, since the low redshift objects require more extrapolation due to their larger apparent extent. We conclude that the high redshift sample which does not contain  any object $M_{500} < \SI{e14}{\textit{h}^{-1}\,M_\odot}$ is most reliable for cosmology.

\subsubsection{Hydrostatic bias}
\label{ch:hydrobias}
Several effects can lead to systematically biased cosmological results, such as instrumental calibrations, substructure, clumping, major merger events or non-thermal pressure which is not accounted in the hydrostatic equation. In \citet{2007ApJ...668....1N} the authors estimate that Chandra mass measurements are biased by 10\% to 20\% low with the respect to the ``true'' masses found in simulations of relaxed clusters. This can originate from subsonic turbulent gas motion (\citealp{evrard}).
With hydrodynamic simulations it is possible to derive hydrostatic equilibrium correction factors for clusters  (e.g., \citealp{2014ApJ...792...25N,2015arXiv150704338S,2015ApJ...808..176A}), but individual clusters might deviate from this trend because of asphericity or clumping. Also the mass accretion rate (i.e. the dynamical state of the cluster) plays an important role. Typically, the unrelaxed clusters show a larger hydrostatic bias.

Observationally there have been studies finding agreement with the predictions on the hydrostatic bias, e.g., by comparing Planck SZ masses derived from XMM-Newton scaling relations with weak-lensing masses (\citealp{2014arXiv1402.2670V}). Other studies find agreement of X-ray masses (mostly from Chandra) with weak-lensing masses (\citealp{2014MNRAS.442.1507G,2015MNRAS.448..814I,2015arXiv150902162A,2015arXiv151101919S}).
\citealp{2015arXiv151107872M} found that X-ray hydrostatic masses are not smaller than masses from galaxy dynamics.
In fact the situation is more complicated since the procedure of estimating a mass matters as well as the weak lensing masses can be biased. Furthermore, in earlier XMM-Newton studies (\citealp{2008A&A...482..451Z,2010ApJ...711.1033Z,2012A&A...546A.106F}) a negligible hydrostatic bias has been found as well. 

\begin{figure}
	\centering
	\resizebox{1.\hsize}{!}{\includegraphics[width=0.49\textwidth]{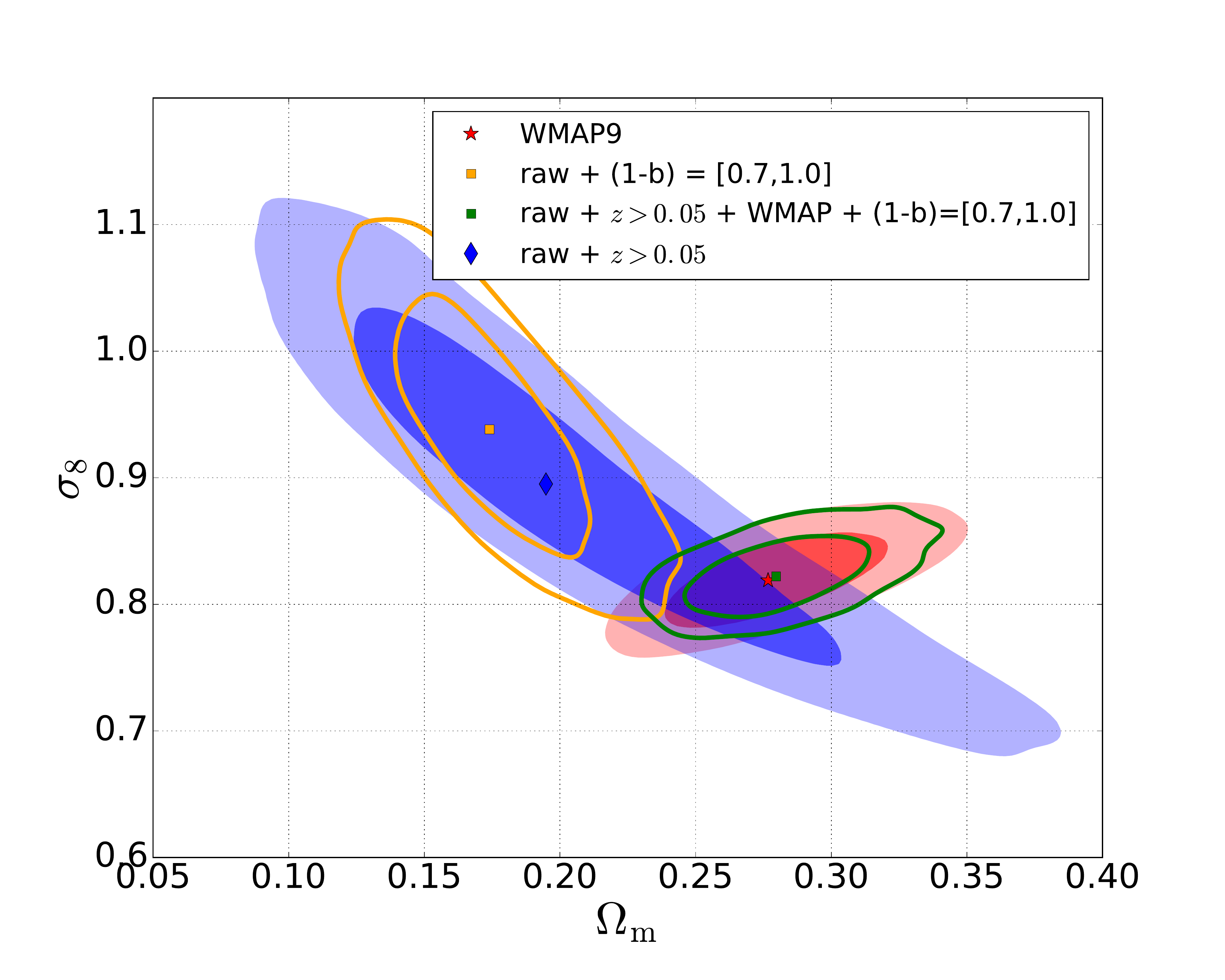}}
	\caption{68.3\% and 95.4\% confidence levels for various setups including a hydrostatic mass bias $(1-b)=\frac{M_\mathrm{hydro}}{M_\mathrm{true}}$. See text for details.}
	\label{fig:bias_1}
\end{figure}

Here we model the mass bias,
\begin{equation}
(1-b) = \frac{M_\mathrm{hydro}}{M_\mathrm{true}}~,
\end{equation}
where both masses are at an overdensity $\Delta = 500$. Furthermore, one can marginalize over a uniformly distributed bias, $(1-b) = [0.7,1.0]$. The shift (see Fig. \ref{fig:bias_1}) is toward higher $\Omega_\mathrm{m}$ and higher $\sigma_8$ values. Also the uncertainties are slightly larger. 

If one uses the large confidence intervals of the high redshift sample from the previous section together with the WMAP9 priors, it is possible to constrain the bias since the $\Omega_\mathrm{m}$ and $\sigma_8$ values are determined almost entirely by WMAP9 (Fig. \ref{fig:bias_1}). The result is $(1-b) = 0.83^{+0.11}_{-0.12}$, which is in agreement with simulations and shows overlap with weak lensing studies comparing their masses to Planck SZ masses, like \textit{Weighing the Giants} ($1-b = 0.70\pm 0.06$ ; \citealp{2014arXiv1402.2670V}) or \textit{CCCP} ($1-b = 0.76 \pm 0.05$ with only statistical uncertainties; \citealp{2015MNRAS.449..685H}). This is a different way of predicting a hydrostatic bias than illustrated before, where masses have been compared to reference values (weak or strong lensing), since now cosmological constraints are compared to a reference (CMB). Since many effects can bias cosmological constraints, this should not be seen as a solid determination of the bias, but as one way of interpreting the HIFLUGCS cosmology results. 

Recent results from simulations by \cite{2016ApJ...827..112B} quantify the hydrostatic bias for various types of clusters (CC or NCC, regular and disturbed) and at several cluster radii ($\Delta = 2500$, $500$, $200$). The authors find that at $\Delta = 500$ the differences for $(1-b)$ between different types of clusters are not significant, so one could adopt a general bias of \mbox{$(1-b)_{\Delta = 500} = \num{0.877(15)}$}. At inner radii, differences are more significant, especially for CC clusters, in which the authors find no significant hydrostatic bias. So we use for our default procedure of the mass function analysis (at $\Delta = 500$) a Gaussian distributed hydrostatic bias, $(1-b) = \num{0.877(15)}$, while for the gas mass fraction analysis (at $\Delta = 2500$) we use $(1-b) = 0.999(27)$ for CC clusters and $(1-b) = \num{0.877(11)}$ for NCC clusters.

\subsubsection{Relaxed clusters}
Many studies require dynamically relaxed clusters (i.e., shape close to spherical, no substructure or major merger) to calibrate their scaling relations. This may have an influence on the cosmological result for the full sample.

Up to now we have used the pure flux cut which creates a sample of relaxed and unrelaxed objects. Unfortunately, there is no general criterion to define relaxed clusters: \citet{2009ApJ...692.1033V} classify clusters with a second emission maximum, filamentary structure or significant centroid shifts as unrelaxed, while \citet{hudson_what_2009} call clusters with round or elliptical isophote and the emission peak in the center of all isophotes relaxed.
\citet{zhang_hiflugcs:_2010} find that clusters with a large offset between the BCG position and the X-ray emission weighted center are often disturbed.
A clearer distinction of clusters can be made via the central cooling time, which splits clusters into cool core or non cool core clusters\footnote{One can also split clusters into three classes, strong-, weak-, and non-cool core clusters as demonstrated in \citet{hudson_what_2009}}. Although the cool core criterion seems to be more objective, there are several disturbed clusters with short cooling times (see \citealp[Fig. 19]{hudson_what_2009}).
We decide to take the disturbed-undisturbed classification from Tab. 2 in \citet{zhang_hiflugcs:_2010}, which is based on the visual inspection of the X-ray flux images as also done in \citet{2009ApJ...692.1033V}. Disturbed clusters are: A0119, A0399, A0400, A0754, A1367, A1644, A1656, A1736, A2065, A2147, A2163, A2255, A2256, A3266, A3376, A3395, A3526, A3558, A3667, MKW8, NGC507. Note that the study by \citet{zhang_hiflugcs:_2010} misses 2A0355 and RXCJ1504, but both of them are strong cool core clusters (\citealp{hudson_what_2009}), so we assume that they are relaxed as well.

\begin{figure}
	\centering
	\resizebox{1.\hsize}{!}{\includegraphics[width=0.49\textwidth]{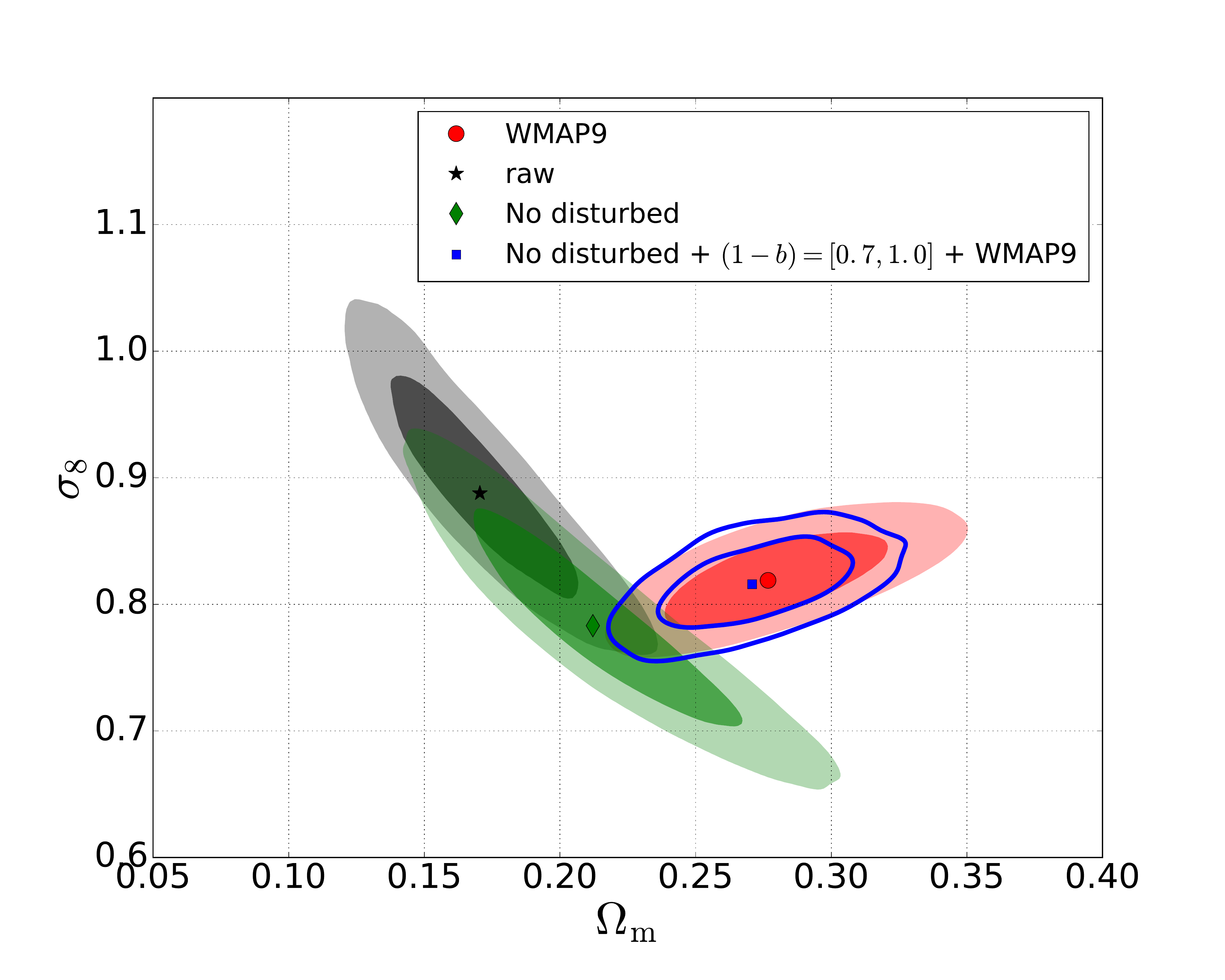}}
	\caption{68.3\% and 95.4\% confidence levels for various setups to demonstrate the effect of dynamically disturbed clusters (raw + undisturbed  in Tab. \ref{tab:cosmo_res2}).}
	\label{fig:disturbed}
\end{figure}

With unconstrained masses for the unrelaxed clusters (Fig. \ref{fig:disturbed}) (i.e. very large errorbars mean that these clusters have small weightings and the mass is constrained mainly by the $L_x-M$ relation which itself is determined by the remaining, relaxed clusters) the constraints of $\Omega_\mathrm{m}$ are shifted toward larger values along the degeneracy with $\sigma_8$.
Adding a hydrostatic bias on this shifts the contours toward higher $\Omega_\mathrm{m}$ and $\sigma_8$ (as shown in the previous section), which will exhibit broad overlap with WMAP9. So we put a uniform prior on the hydrostatic bias and added WMAP9 priors on the two variable cosmological parameters. The resulting bias (Fig. \ref{fig:disturbed}) is $(1-b) =0.66^{+0.10}_{-0.12}$, which slightly overlaps with the 68.3\% uncertainties of the bias constraints from the high redshift subsample combined with WMAP9. 

\subsection{Independent mass estimates}
In Paper I, we compared our hydrostatic masses to the available Planck SZ masses as well as to the dynamical masses from optical velocity dispersions using the caustic method to select member galaxies. 
We show now briefly a comparison of the cosmological parameters using these alternative masses for (part of) the HIFLUGCS clusters. 

\begin{figure*}
	\centering
	\resizebox{1.\hsize}{!}{\includegraphics[width=0.95\textwidth]{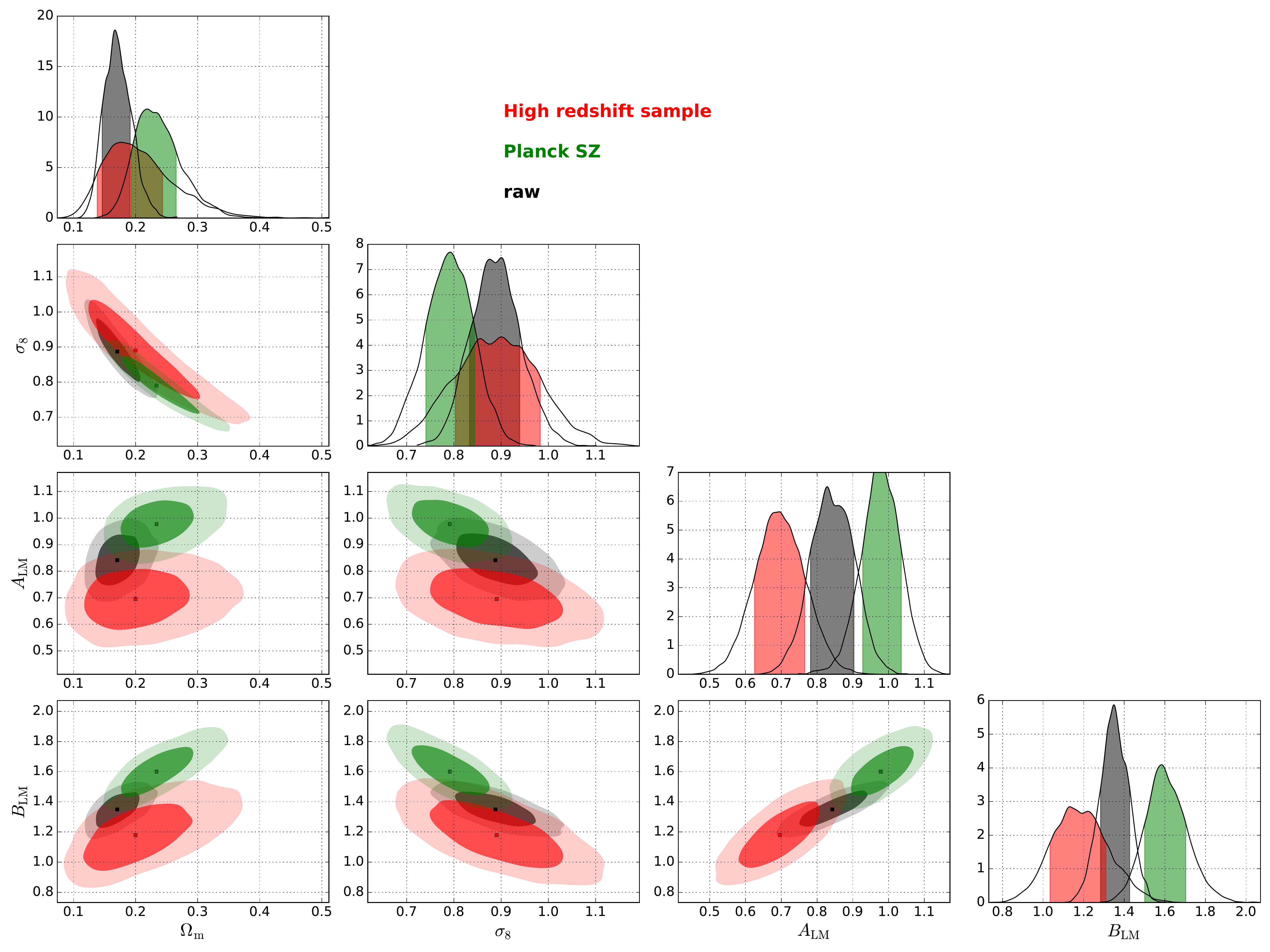}}
	\caption{68.3\% and 95.4\% confidence levels for the analysis of the Planck SZ masses of HIFLUGCS (green), compared to the raw analysis procedure (black) and the high redshift subsample (raw + $z>0.05$, red).}
	\label{fig:planckmasses}
\end{figure*}

Figure \ref{fig:planckmasses} shows that for the Planck SZ masses of HIFLUGCS, there exists an overlap for $\Omega_\mathrm{m}$ and $\sigma_8$ with the high redshift subsample of the raw analysis procedure, as expected since 12 low redshift clusters are already excluded. The $L_x-M$ relation constrained from the Planck masses (using the same luminosities) deviates strongly from the default or high redshift sample: The slope and normalization are about 30\% larger than the high redshift subsample constraints. A higher normalization of the $L_x-M$ relation could be due to a systematic bias that lowers all Planck masses. Following the degeneracy between slope and normalization, the slope has to increase as well. The Chandra$-$XMM-Newton cross calibration could contribute to this effect, but it is unlikely that it is fully attributable to calibration uncertainties, which would not cause such a strong deviation. In Paper I we discuss the differences between our hydrostatic and Planck masses in more detail.

We use the mass estimates by \cite{2016arXiv160806585Z} based on the velocity dispersion of the HIFLUGCS sample from \citet{zhang_hiflugcs:_2010} with some updates. 
\begin{figure*}
	\centering
	\resizebox{1.\hsize}{!}{\includegraphics[width=0.95\textwidth]{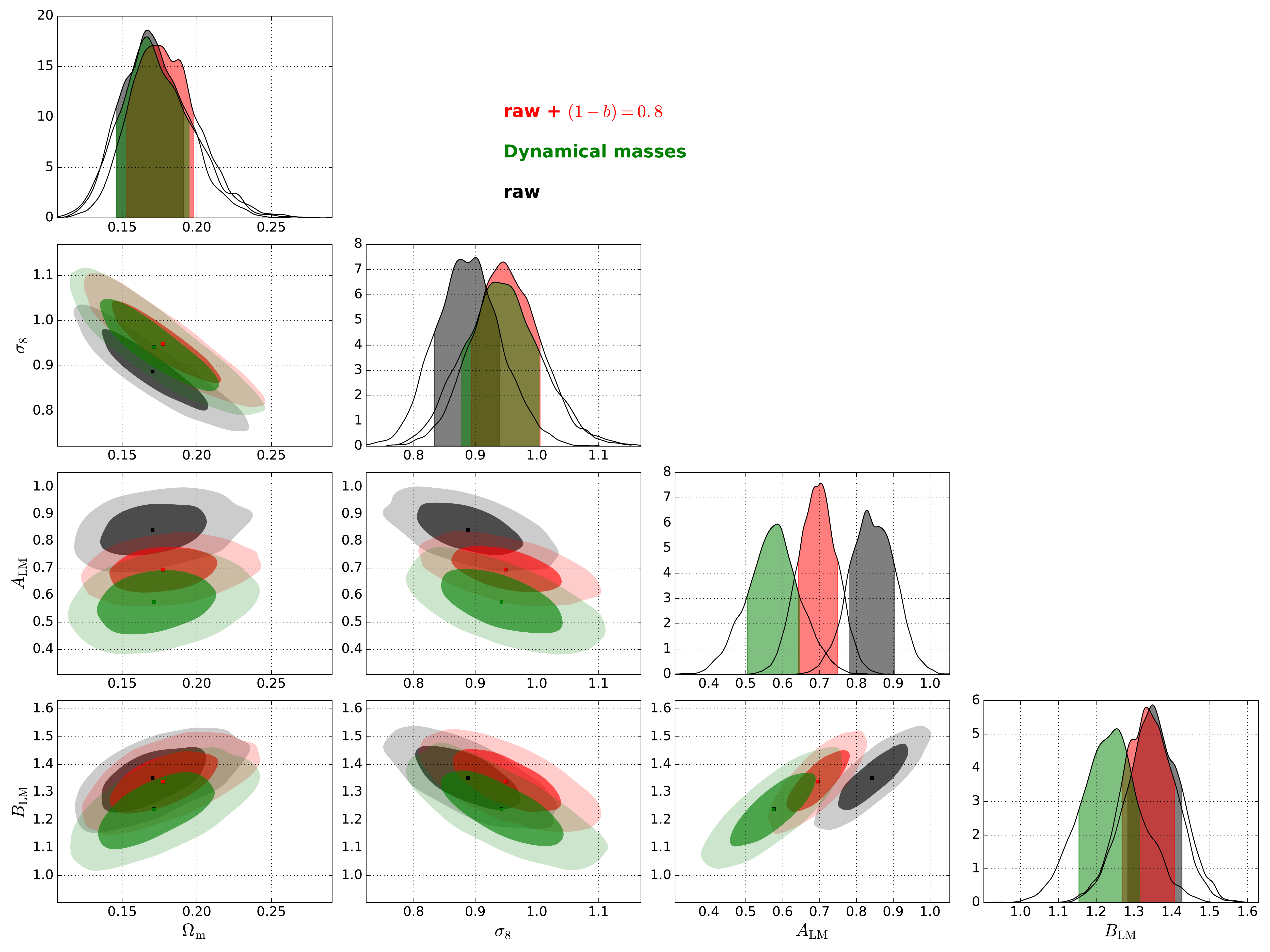}}
	\caption{68.3\% and 95.4\% confidence levels for the analysis of the dynamic masses of HIFLUGCS (green; \citealp{2016arXiv160806585Z}), compared to the raw analysis procedure (black) and the raw setup plus a fixed hydrostatic bias of $(1-b)=0.8$ (red).}
	\label{fig:dynmasses}
\end{figure*}
The two missing clusters (2A0355 and RXCJ1504) have very large uncertainties so their mass is determined by the $L_x-M$ relation. The direct comparison to the (raw) hydrostatic masses in Paper I shows that despite the scatter the dynamic and hydrostatic masses are in good agreement, so it is no surprise the cosmological results are also matching (Fig. \ref{fig:dynmasses}). As indicated in Fig. \ref{fig:dynmasses} adding a hydrostatic bias $(1-b) = 0.8$ to the default masses gives even better agreement with the dynamic mass cosmology. This independent mass estimator shows that the hydrostatic masses derived here are robust: First, there is no mass dependency between the Chandra derived hydrostatic masses and the dynamic mass estimates, which can give a hint that the Chandra instrumental calibration is reliable. Second, the offset of the mass-mass comparison (see Paper I) is consistent with zero ($-0.02 \pm 0.04$), which also makes the existence of a large amount of unaccounted non-thermal pressure less likely. 

\subsection{Influence of cosmic variance}
\label{ch:cosmic_variance}
\begin{figure}
	\centering
	\resizebox{1.\hsize}{!}{\includegraphics[width=0.5\textwidth]{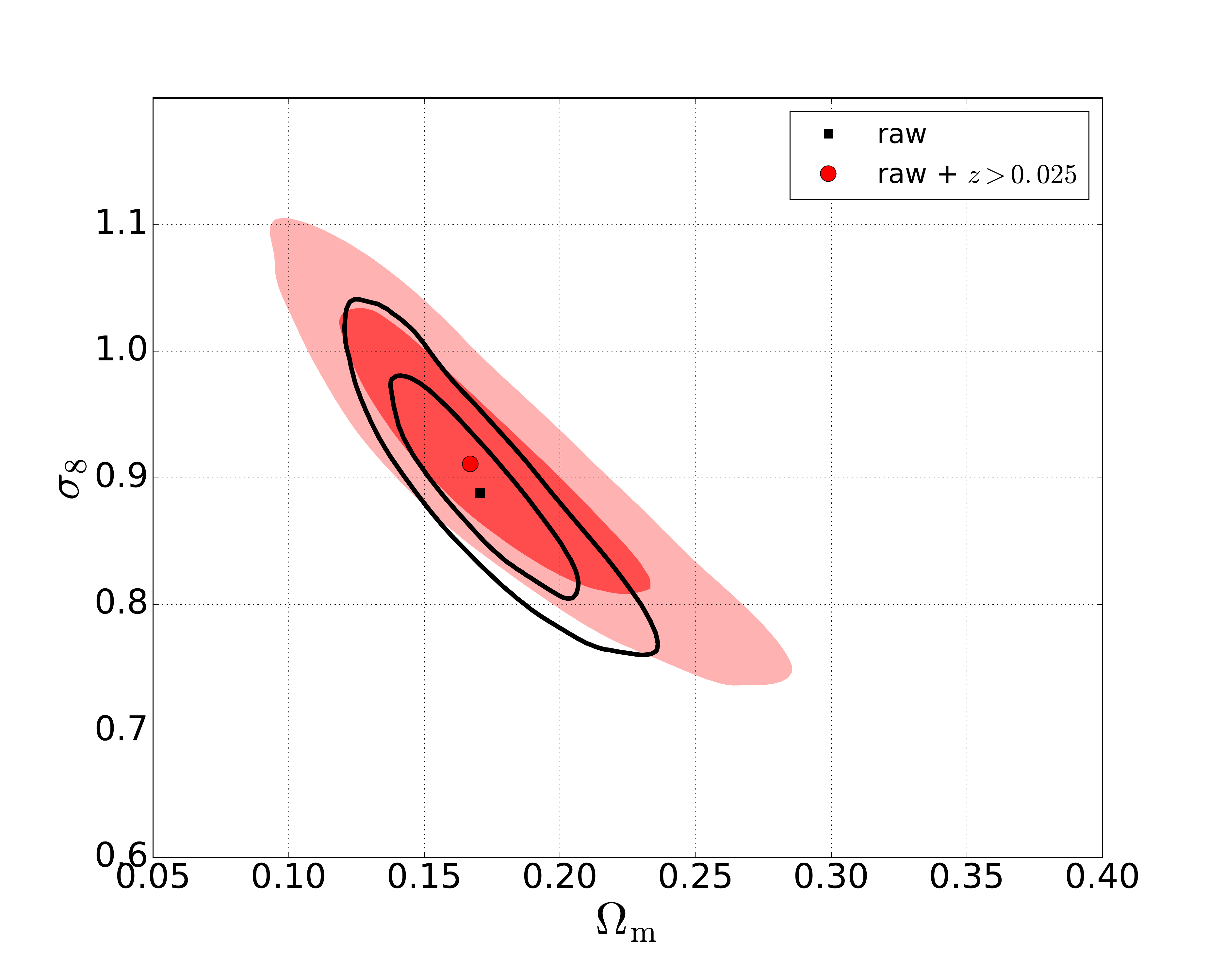}}
	\caption{Influence of cosmic variance on the $\Omega_\mathrm{M}$ - $\sigma_8$ determination. The red confidence intervals excludes the 13 clusters within $\SI{100}{Mpc}$.}
	\label{fig:cosmic_variance}
\end{figure}

The number of galaxy clusters within a certain volume emerges from the fluctuations of the large-scale matter density distribution. 
Especially within small volumes the statistical fluctuations related to the sample size dominate over other sources of uncertainty.
For HIFLUGCS, the most nearby galaxy clusters populate a relatively small volume and their number density could be highly biased in either direction. At the same time, these nearby objects are also low mass galaxy groups, and make an important contribution to the halo mass function. To test the influence of cosmic variance on our cosmological results, we exclude the 13 objects within $\SI{100}{Mpc}$. If our local Universe was biased in either direction, one should be able to see a significant shift of the cosmological parameters. Figure \ref{fig:cosmic_variance} shows that no influence of the 13 closest objects can be detected, so unless the Universe is inhomogeneous on larger scales than expected, we conclude that cosmic variance has a negligible influence on our results.

\subsection{Reference model validity}
\label{ch:hicosmo_model}
In this final part of the discussion on the HIFLUGCS cosmology results of the mass function analysis we focus on the mass function itself. For the construction of the \citet{2008ApJ...688..709T} halo mass function Dark Matter simulations without baryons were used. We point out that the following tests are used to demonstrate the impact of these components, but not to put any constraints on them. This is not easily possible with galaxy clusters alone, especially not with a local sample.

\subsubsection{Impact of baryons}
The clear difference between baryonic matter and Dark Matter is the fact that Dark Matter only interacts gravitationally. Dark Matter is seen as collisionless and so does not have pressure. Obviously the Universe consists of baryonic matter, but due to its interaction simplicity and the much higher energy density in the Universe, simulations in the past were often performed using Dark Matter particles and neglecting baryons.

Up to now we only included the properties of the baryonic matter in the matter power spectrum (or matter transfer function): The interaction of the baryons and the tightly coupled photons in the early Universe led to oscillations (Baryonic acoustic oscillations, e.g., \citealp{2005NewAR..49..360E}). These are imprinted on the matter power spectrum, since baryons change the total gravitational potential. Apart from the oscillations a damping can also be observed due to photons smoothing out the small scale temperature fluctuations at the epoch of recombination. More details are given in \citet{1998ApJ...496..605E,2005NewAR..49..360E}.  The effect of baryons on the transfer function is fully accounted for by the CLASS software.

For the mass function the situation is the following: By adiabatic contraction baryons force a larger halo concentration, which leads to a uniform shift of halo masses toward larger values (e.g., \citealp{2012MNRAS.423.2279C}). Furthermore, baryons cause several physical effects that are accounted in current hydrodynamic simulations (\citealp{2014MNRAS.441.1769C,Bocquet2015a}), such as gas heating, radiative cooling, star formation and feedback processes from active galactic nuclei (AGN) and galactic winds driven by supernovae. It was found that the impact of baryons is much larger in the inner regions (e.g., $r_{2500}$) and almost negligible at the virial radius. Cooling and star formation lead to a slightly larger halo density (i.e. increase of the mass function with respect to Dark Matter only derived mass functions), while adding AGN feedback on top can result in a suppression of the mass function (\citealp{2014MNRAS.441.1769C}).
This has also been shown with a different type of simulations (AMR instead of SPH) by \cite{2014MNRAS.440.2290M}, where the authors demonstrate that the effect of baryons on cosmological parameters is on a percent level.

Here we adopt the mass function from \citet{Bocquet2015a}, which includes stellar evolution, chemical enrichment, star formation and feedback processes (AGN and SNe). The simulations are based on three different box sizes of the \textit{Magneticum} simulation with an input cosmology $\Omega_\mathrm{m} = 0.272, \Omega_\mathrm{b} = 0.0456, \sigma_8 = 0.809$ and $h = 0.704$. 
\citet{Bocquet2015a} predict for $z<0.3$ that the mass function from hydrodynamic simulations (including the baryonic effects, HydroMF in the following) will be smaller than \citet{2008ApJ...688..709T} at all masses, while the difference seems to be minimal around $\SI{e15}{M_\odot}$. Naively that would mean that the HydroMF gives larger $\Omega_\mathrm{m}$ than \citet{2008ApJ...688..709T}. Of course the impact will depend on the individual sample selection. For example, for an eROSITA like sample (X-ray flux selected) the  \citet{2008ApJ...688..709T} mass function would predict lower $\Omega_\mathrm{m}$ and smaller $\sigma_8$. Note that the input cosmology for these tests in \citet{Bocquet2015a} was close to the cosmology used in the simulations.

\begin{figure}
	\centering
	\resizebox{1.\hsize}{!}{\includegraphics[width=0.49\textwidth]{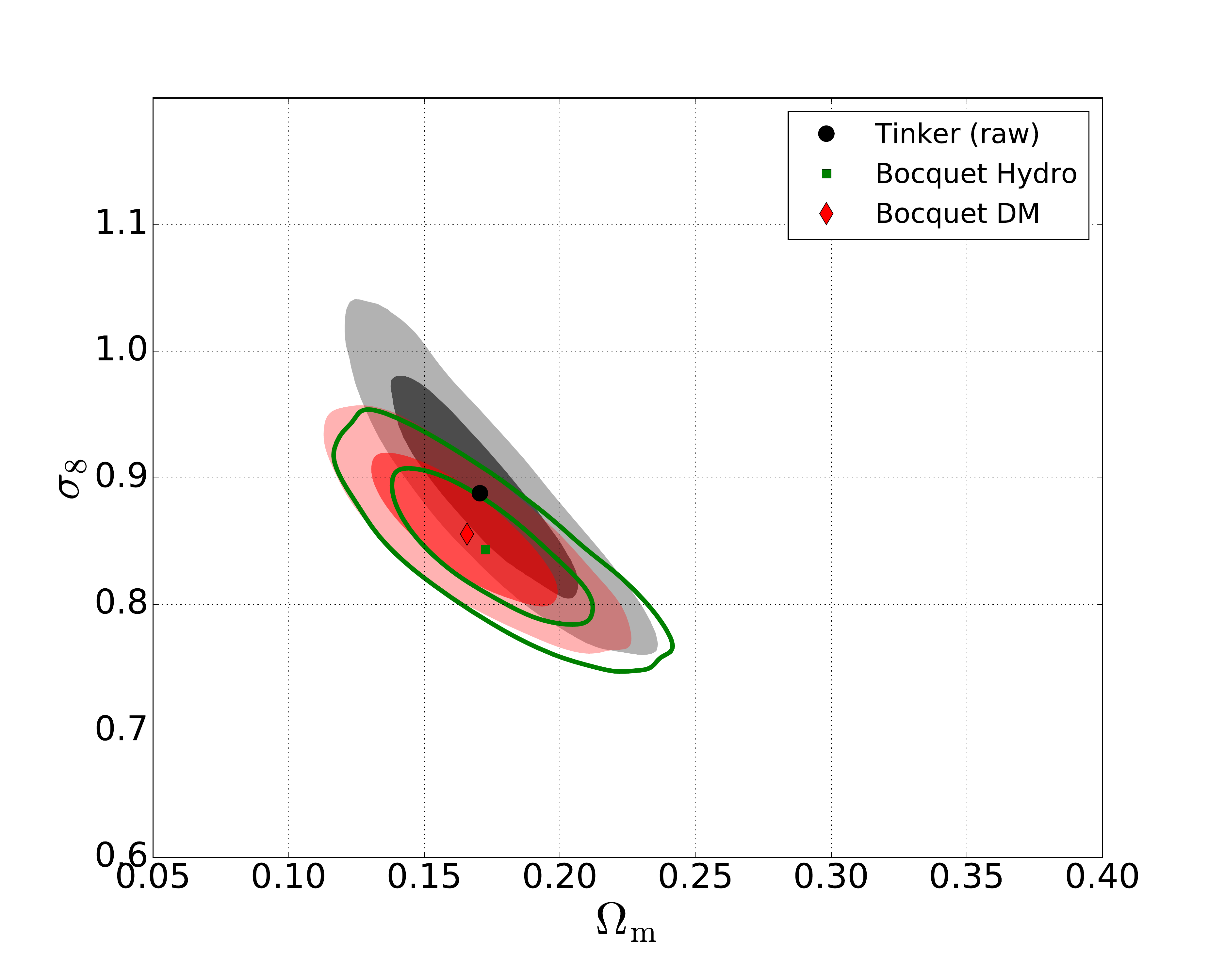}}
	\caption{68.3\% and 95.4\% constraints on $\Omega_\mathrm{m}$ and $\sigma_8$ from the mass functions by \citet{Bocquet2015a}. Green ellipses are from hydrodynamical simulations, while the red ellipses are Dark Matter only simulations. The \citet{2008ApJ...688..709T} constraints (``raw'') are shown in black.} 
	\label{fig:bocquet}
\end{figure}

Applying the \citet{Bocquet2015a} mass function to HIFLUGCS (raw) results in a different trend (see Fig. \ref{fig:bocquet}): We detect no change in $\Omega_\mathrm{m}$ and a lower $\sigma_8$ when using the HydroMF, which is not expected. The Dark Matter only mass function from \citet{Bocquet2015a} shows good agreement with the HydroMF, which is also not expected. The uncertainties of the HydroMF are larger, mostly because the covariance matrix enters and puts uncertainties on the parameters, which was not done for the Dark Matter only mass function of \citet{2008ApJ...688..709T} and \citet{Bocquet2015a}. Directly comparing the HydroMF mass function against \citet{2008ApJ...688..709T} reveals that for $\Omega_\mathrm{m} \approx 0.17$ the situation is the reverse of that described in \citet{Bocquet2015a}: \citet{2008ApJ...688..709T} gives a lower halo density than HydroMF for high masses and shows agreement for low masses, which is locally equivalent to a change in $\sigma_8$ as observed. One reason could be that one (or both) mass functions compared here are not universal at $\Delta_\mathrm{500c}$. As indicated already in \citet{Bocquet2015a} the best universality is given at an overdensity of $\Delta_\mathrm{200m}$. It seems that for HIFLUGCS the effects of baryons in the mass function is negligible (if HydroMF is correct), especially if one compares the tiny difference between the Dark Matter only (orange) and HydroMF (green) constraints of the \citet{Bocquet2015a} mass functions in Fig. \ref{fig:bocquet}. The differences between different Dark Matter only simulations on the other hand can be significant, as in the described case there is only small overlap of the 68.3\% confidence regions.

\subsubsection{Neutrinos}
\label{ch:neutrinos}
In the standard model of particle physics neutrinos are considered as massless particles without charge, although several studies have concluded that they have a non-zero mass (e.g., \citealp{1999ARNPS..49...77K,2004NJPh....6...76H,2012PhRvD..86a3012F,2013AstL...39..357B}). 
A massive neutrino will have an effect on the large scale structure that can be seen in the matter power spectrum (e.g., \citealp{2006PhR...429..307L,2012arXiv1212.6154L,2014NJPh...16f5002L}): On sub-free-streaming-scales neutrinos will smooth the density field, which makes the gravitational potential shallower and slows down structure growth (\citealp{2011MNRAS.410.1647A}). 
The distribution of masses among the neutrino species is uncertain, but the sum of the masses of the three species, $\sum m_\nu$, is important for our analysis. 

The neutrino mass enters in the calculation of the matter power spectrum and it is taken into account by CLASS (\citealp{2011arXiv1104.2932L,2011arXiv1104.2934L,2011JCAP...09..032L}).
Massive neutrinos are no longer in the relativistic regime, but can be described by a Fermi-Dirac distribution. The mass can be translated into an energy density,
\begin{equation}
\Omega_\nu = \frac{\sum m_\nu}{\SI{93.14}{eV}\,h^2}~,
\end{equation}
which is taken into account when calculating the total matter density. 
The oscillation experiments now determine three neutrino mass states, and each mass corresponds to a superposition of the three different neutrino flavors. Theory predicts two different scenarios for the distribution of mass states: The normal and inverse hierarchy (see also \citealp{2013arXiv1307.5487C}), with each having different implications on the standard model of particle physics. However, for cosmological applications, the sum of neutrino masses is important.

\begin{figure}
	\centering
	\resizebox{1.\hsize}{!}{\includegraphics[width=0.49\textwidth]{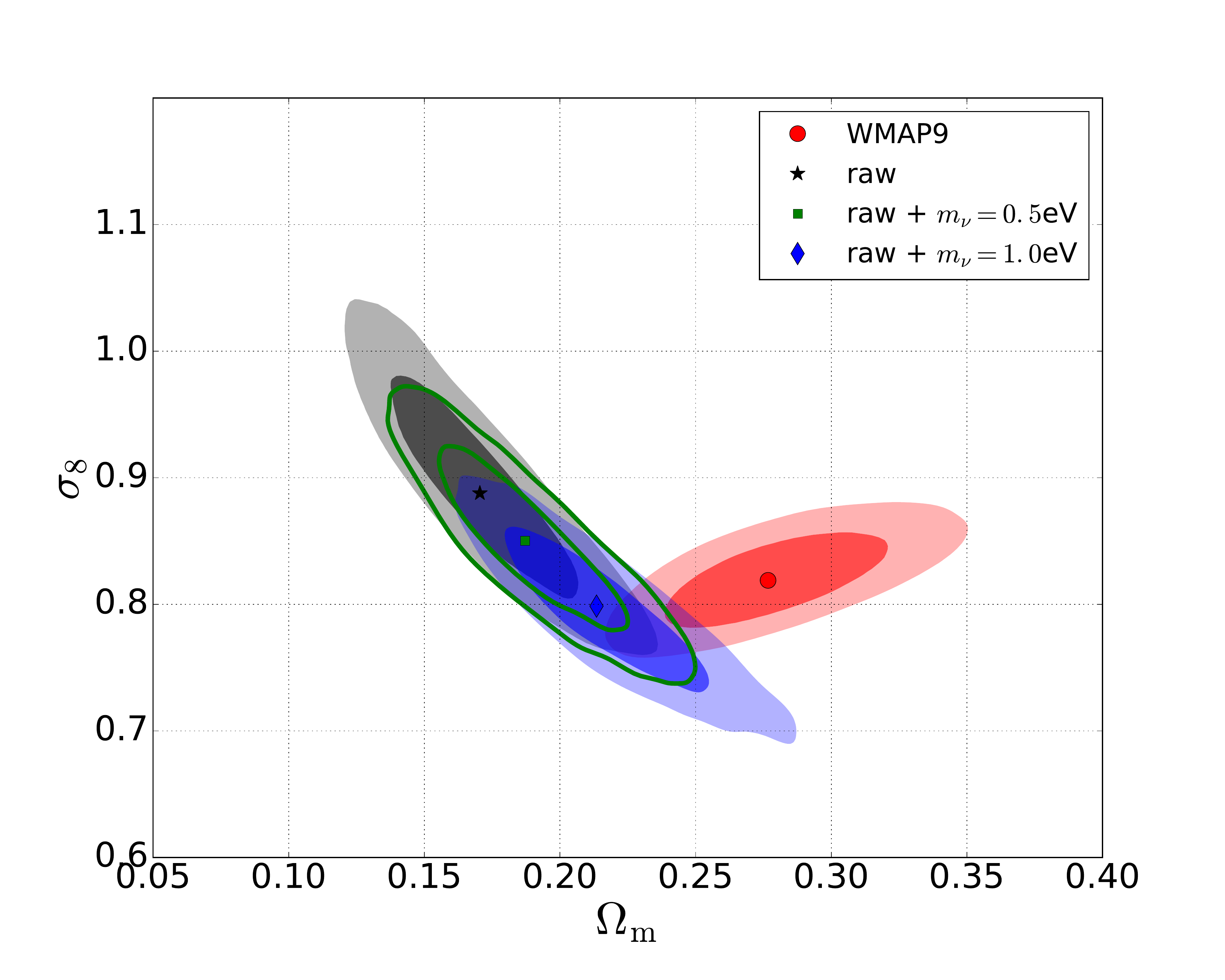}}
	\caption{68.3\% and 95.4\% constraints on $\Omega_\mathrm{m}$ and $\sigma_8$ for an increased (summed) mass of neutrinos (raw: $\sum m_\nu=\SI{0.06}{eV}$).} 
	\label{fig:neutrino}
\end{figure}

The results for the REFLEX2 galaxy cluster sample (\citealp{2015A&A...574L...8B}) suggest a smaller $\sigma_8$ and larger $\Omega_\mathrm{m}$, when increasing the neutrino mass. This is in good agreement with what is obtained for HIFLUGCS (see Fig. \ref{fig:neutrino}). 
The cases shown here ($\sum m_\nu$ up to $\SI{1}{eV}$) are unrealistically high. 
Current upper limits are found by \cite{2015arXiv150201589P} ($\SI{0.17}{eV}$), and by \cite{2017arXiv170108172V} ($\SI{0.12}{eV}$), both at 95\% confidence level,
but our examples are shown here to demonstrate the mass at which the effect would be significant.

The constraints on the $L_x-M$ relation are unchanged for any high neutrino mass tested here. The effect in the $\Omega_\mathrm{m}$-$\sigma_8$ plane is perpendicular to the hydrostatic bias, 
which makes it possible to approach the WMAP9 results by
adding a sufficiently high bias and neutrino mass, which both would be unphysical. As stated before, it is not our aim to reproduce CMB results.

\subsection{Gas mass measurements}
\begin{figure}
	\centering
	\resizebox{1.\hsize}{!}{\includegraphics[width=0.49\textwidth]{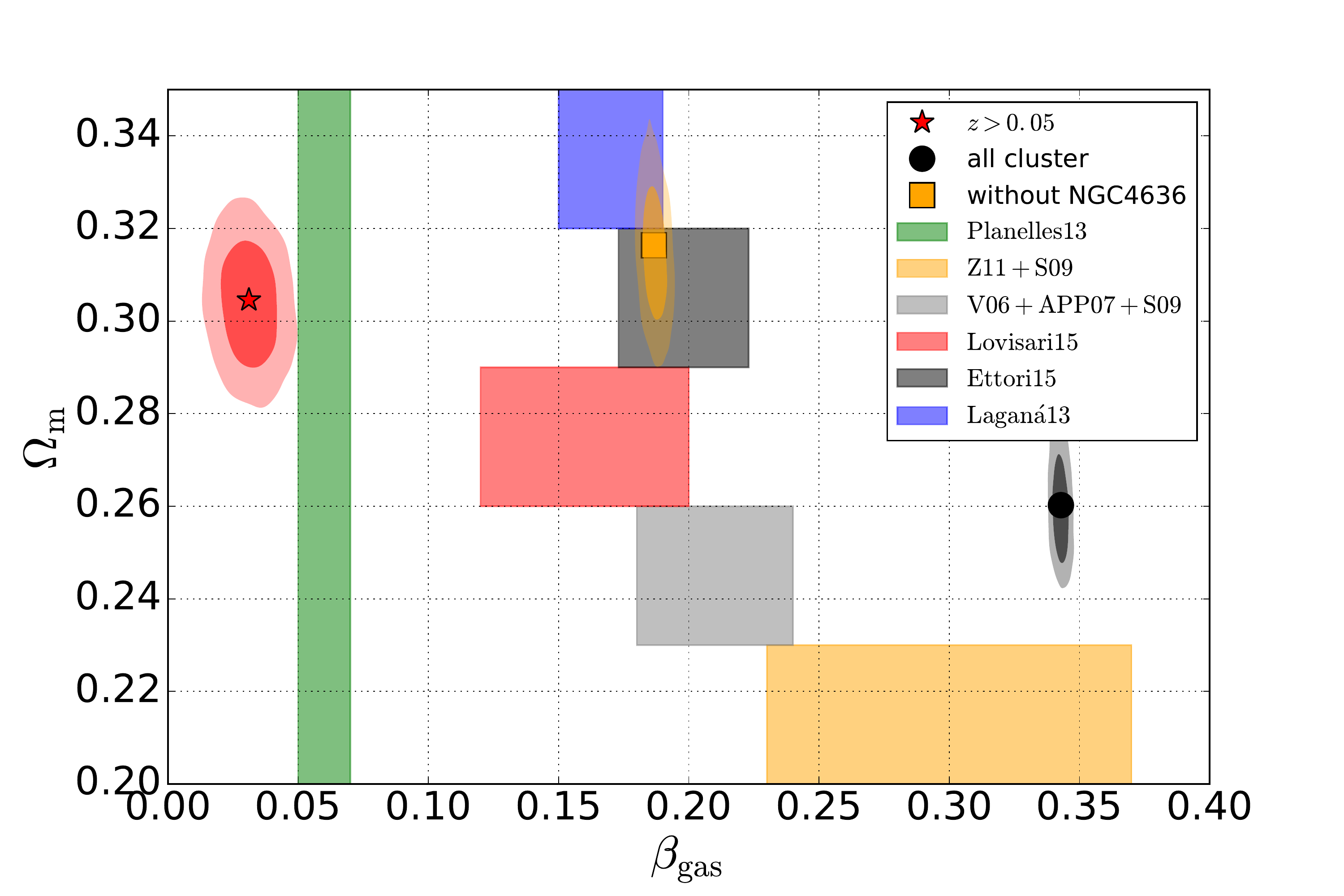}}
	\caption{68.3\% and 95.4\% confidence levels of the $f_\mathrm{gas}$ tests for $\Omega_\mathrm{m}$ and $\beta_\mathrm{gas}$, the mass dependence of the gas depletion factor, for the HIFLUGCS sample, calculated at $r_{2500}$. See text for details.}
	\label{fig:fgas_cosmo3}
\end{figure}

The results of the $f_\mathrm{gas}$ test (Section \ref{ch:gasmass_result}) alone can only constrain $\Omega_\mathrm{m}$, but not $\sigma_8$. The derived constraints on $\Omega_\mathrm{m}$ of the $z>0.05$-sample can be added as a prior on the halo mass function analysis. As shown in Fig. \ref{fig:ellipse_def} this additional information eliminates the degeneracy between $\Omega_\mathrm{m}$ and $\sigma_8$. 
For the setup including the $f_\mathrm{gas}$ priors one cannot detect any significant difference of the $L_x-M$ relation compared to the MF-only setup. 
The slightly different $\Omega_\mathrm{m}$ constraints of these two analysis procedures (halo mass function and $f_\mathrm{gas}$ test) of the same sample can probably be explained by the sensitivities of the test: $f_\mathrm{gas}$ is compared to hydrodynamic simulations with many degrees of freedom in the fit. 
In particular, the mass dependence modeled by $\beta_\mathrm{gas}$ in Eq. \ref{eq:fgasmodel2} gives flexibility to account for the different behavior of galaxy groups. The values of $\beta_\mathrm{gas}$ in Table \ref{tab:fgas1} and Fig. \ref{fig:fgas_cosmo3} show that simulations do not reproduce the observed properties of galaxy groups well (high values of $\beta_\mathrm{gas}$), but since this parameter is free, the fit is able to account for the observed lower gas mass fractions in groups (which is also supported by other studies).
As found by \citet{2006ApJ...640..691V,2007A&A...474L..37A,2009ApJ...693.1142S} (gray), \citet{2009ApJ...693.1142S,2011A&A...535A..78Z} (yellow), \citet{Lovisari2015} (red), \citet{2015MNRAS.446.2629E} (black) or \citet{2013A&A...555A..66L} (blue), all shown in Fig. \ref{fig:fgas_cosmo3}, the mass dependence of the gas depletion factor is typically around 0.15 to 0.25, but depends on the sample composition. In our sample there are several low mass galaxy groups. Especially the lowest mass system, NGC4636, seems to introduce a large bias for $\beta_\mathrm{gas}$: If one excludes only this group from the $f_\mathrm{gas}$ analysis, the value of $\beta_\mathrm{gas}$ is reduced by almost 50\% (Fig. \ref{fig:fgas_cosmo3}), making it consistent with the other studies mentioned before. Note that the colorbars in Fig. \ref{fig:fgas_cosmo3} should not refer to a specific value or range of $\Omega_\mathrm{m}$, but are just placed in the figure for comparison reasons.
Since neither the simulations, nor other studies have included galaxy groups with this low mass in their studies, the high value of $\beta_\mathrm{gas}$ could be real. 
But conclusions based on just one single object are not justified.

\section{Conclusion}
We present a cosmological analysis using a complete, X-ray selected, purely flux limited sample of 64 local galaxy clusters and calibrate the $L_x-M$ scaling relation (Eq. \ref{eq:lxm}), for whose parameters we find $A_\mathrm{LM} = \num{0.67(7)}$ and $B_\mathrm{LM} = 1.34^{+0.13}_{-0.12}$.
Masses of the galaxy clusters have been calculated individually for each cluster from the temperature and surface brightness profiles. A crucial step was to perform the extrapolation of the mass, since the Chandra FOV and, for some clusters, the limited exposure time, does not allow us to measure the temperature and its gradient at $R_{500}$. Several methods are applied to perform the extrapolation, either simply by using models for the temperature profiles, which can produce (unphysical) decreasing total mass profiles in the outer regions, or by using an NFW model (see Paper I, for more details). The NFW fit with a concentration parameter linked to the total mass was used as the default. Additionally we noticed with the $R_{500}$-test (using a subsample of clusters whose temperature profile reaches $R_{500}$) that cluster masses are underestimated depending on the extrapolation to reach $R_{500}$. Although for most clusters this is a small effect, we took this into account.

We performed several tests to measure the robustness of these constraints: Simple tests like excluding some extreme clusters (fluxes which are just at the flux limit, large deviations from the $L_x-M$ relation), did not show a significant effect. 
We also confirmed past results indicating that the Chandra/XMM-Newton cross calibration has no significant effect on the cosmological parameters for the HIFLUGCS sample.
Systematic modifications of the sample selection function, like a redshift cut, revealed an interesting trend: The high redshift subsample shows larger uncertainties, as expected, but is also in better agreement with CMB measurements. The low redshift subsample, containing all the groups of the full sample, exhibits a much steeper $L_x-M$ relation and even lower $\Omega_\mathrm{m}$ values. Since some effects, like the influence of galaxy groups, seems to be important, this was evaluated in more detail: 

\begin{itemize}
	\item Galaxy groups: The parent sample may have a higher incompleteness and/or missing flux fraction at the galaxy group scale. This could be corrected either by reanalyzing the RASS data with a source detection algorithm adapted to special group properties (e.g., lower surface brightness or flatter surface brightness profiles), or by quantifying their influence, i.e., assuming a higher incompleteness for groups (see Section \ref{ch:groups}). The latter test revealed that for our sample $\Omega_\mathrm{m}$ might be biased low by about $0.03$ in the raw analysis. Furthermore, the astrophysical processes in galaxy groups might be different, which results in a steep $L_x-M$ relation for low mass objects and the powerlaw might not be a good parametrization. This can only be better described with large, complete samples of galaxy groups, where statistical uncertainties are small and selection effects can be corrected for.
	\item Hydrostatic mass bias: Simulations and comparisons of the X-ray and weak lensing masses of galaxy clusters point to the existence of a hydrostatic mass bias. Using alternative cluster masses (e.g., dynamical masses from velocity dispersion) we find very good agreement if we assume a bias of $(1-b) = 0.8$ for the X-ray masses. Dynamically disturbed clusters, which often violate (to some extent) hydrostatic equilibrium, can bias results, since the exclusion of this subsample leads to higher values of $\Omega_\mathrm{m}$ and lower values of $\sigma_8$. But assuming that reference cosmological results (e.g., WMAP9) are correct, one can obtain an indirect mass calibration when leaving the hydrostatic bias free to vary. For our default analysis we rely on the very recent simulation results by \cite{2016ApJ...827..112B}.
	\item Extrapolation: For many of our clusters an extrapolation is needed to constrain the total mass at $R_{500}$. The only way to avoid this is to gather many expensive observations at large cluster radii. For our analysis, we correct our NFW extrapolation model with a few clusters having temperature measurements to large radii. Although for most clusters this correction is small, we find a systematic shift of $\Omega_\mathrm{m}$ by $0.01$ due to this correction.
	\item Theoretical mass function: Comparing two different halo mass functions based on N-body simulations results only in a small shift of $\sigma_8$ by $\sim 0.03$. 
	Baryonic effects (e.g., feedback processes) are expected to play a significant role for low mass systems. Using a mass function which incorporates these effects shows no changes: Compared to a consistently derived Dark Matter only mass function, the change is completely insignificant.
	\item Other physical/cosmological effects: Several physical and/or cosmological effects could help (e.g., here we show explicitly the effects of neutrinos with masses $\gg\SI{0.06}{eV}$), but the other effects above need to be controlled better before robust conclusions can be drawn.	
\end{itemize}

In sum, the effect of galaxy groups and/or disturbed objects (in the raw analysis) in HIFLUGCS seems to bias results toward lower $\Omega_\mathrm{m}$. 
Due to large uncertainties in the results from subsamples, the changes are often insignificant.
The $f_\mathrm{gas}$ test also shows that galaxy groups need to be treated more carefully. 
Simulations are not able to recover the values of $f_\mathrm{gas}$ for low mass system, raising the suspicion that halo mass function simulations may also not reproduce the number density of those systems well.

The final cosmological results with purely statistical (68\%) uncertainties from the halo mass function (\citealp{2008ApJ...688..709T}) give $\Omega_\mathrm{m} = 0.22^{+0.07}_{-0.05}$ and $\sigma_8  = \num{0.89(10)}$, and after adding the prior from the gas mass fraction analysis we get our default results: $\Omega_\mathrm{m} = \num{0.30(1)}$ and $\sigma_8 = \num{0.79(3)}$. 
These include corrections and modifications based on the tests presented in the discussion: The effects of galaxy groups and the uncertainties of cosmic variance in the very nearby Universe (leading to a lower redshift cut of $z>0.05$), and the corrections for the mass extrapolation and hydrostatic mass biases. 

Including these results for systematics in a broader framework, e.g., by increasing the sample size (\mbox{eHIFLUGCS}), will enable us to estimate more detailed cosmological parameters (the Dark Energy density, $\Omega_\mathrm{DE}$, and its equation of state).

\section*{Acknowledgements}
The authors would like to thank the anonymous referee for constructive feedback, and Ewan O'Sullivan for providing helpful comments. GS, THR acknowledge support by the German Research Association (DFG) through grant RE 1462/6. GS acknowledges support by the Bonn-Cologne Graduate School of Physics and Astronomy (BCGS) and the International Max Planck Research School (IMPRS) for Astronomy and Astrophysics at the Universities of Bonn and Cologne. THR acknowledges support by the DFG through Heisenberg grant RE 1462/5 and the Transregio 33 ``The Dark Universe'' sub-project B18.
This research made use of Astropy, a community-developed core Python package for Astronomy.


\bibliographystyle{mnras}
\bibliography{Astro}

\appendix
\onecolumn

\section{Likelihood function}
\label{ap:likelihood}
In order to get rid of biases arising from binning the clusters in mass, flux, luminosity and/or redshift, one can do the transition to make these volume bins as small as possible, as done, e.g., by \citet{2008MNRAS.387.1179M,2009ApJ...692.1033V}. A corresponding likelihood function would then comprise a simple source counting within a Bayesian regression model:
\begin{equation}
\label{eq:likelihood1}
\mathcal{L}(\hat{M},\hat{L},N_\mathrm{det}) = \underbrace{\frac{\langle N \rangle^N \, e^{- \langle N\rangle}}{N!}}_\mathrm{Poisson} \cdot \underbrace{\frac{N!}{N_\mathrm{det}! \, N_\mathrm{mis}!}\, p^{N_\mathrm{det}} \, \left( 1-p \right)^{N_\mathrm{mis}}}_\mathrm{Binomial} \cdot \underbrace{\prod \limits_{i=1}^{N_\mathrm{det}} \tilde{P_i}}_{\mathrm{observational\,probability}} ~.
\end{equation}
The hat, $\hat{x}$, on parameters marks observed quantities.
The first part is the Poisson likelihood for predicting $\langle N \rangle$ clusters with the halo mass function (without selection) while having $N$ clusters in the Universe. The second term is a Binomial likelihood for detecting $N_\mathrm{det}$ clusters using the selection function out of the $N$ clusters and missing $N_\mathrm{mis} = N - N_\mathrm{det}$ clusters in the sample. $p$ is the probability for all the detected sources to be detected within the current constraints on the cosmology, scaling relation and selection function, while $(1-p)$ is the probability to miss $N_\mathrm{mis}$ sources. 
The last term is the probability for observing each individual cluster with its properties like mass and luminosity, so $\tilde{P}_i$ depends on the $\hat{M}$, $\hat{L}$, the parameters of the scaling relation and the selection criteria.
Following the derivations in \citet{2010MNRAS.406.1759M}, one can rewrite the parameters in the following:
\begin{equation}
\langle N \rangle = \int \mathrm{d}z\,  \frac{\mathrm{d}V}{\mathrm{d}z}\cdot  \int \mathrm{d}M\, \Omega_\mathrm{frac}(z,M) \cdot \frac{\mathrm{d}n}{\mathrm{d}M} ~,
\end{equation}
where $\langle N \rangle$, as mentioned before, is the total number of predicted clusters and
\begin{equation}
\label{eq:likelihood4}
\langle N_\mathrm{det} \rangle = \int \mathrm{d}z\,  \frac{\mathrm{d}V}{\mathrm{d}z}  \int \mathrm{d}M \, \Omega_\mathrm{frac}(z,M) \cdot \frac{\mathrm{d}n}{\mathrm{d}M} \int \mathrm{d}L\, \int \mathrm{d}\hat{L} \cdot \mathcal{N}(L,\hat{L},\sigma_{\hat{L}}) \cdot \mathcal{N}(L,L_\mathrm{LM},\sigma_\mathrm{LM}) \cdot P_I~,
\end{equation}
is the number of predicted clusters accounting for the selection function, where $\frac{\mathrm{d}V}{\mathrm{d}z}$ is the comoving volume element at redshift $z$, $\Omega_\mathrm{frac}(z,M)$ is the covered sky fraction (which is assumed to be constant for HIFLUGCS, but will be tested to depend on redshift or mass) and $\frac{\mathrm{d}n}{\mathrm{d}M}$ is the halo mass function (halo number density per mass, e.g., by \citealp{2008ApJ...688..709T}). The fact that the sky fraction can depend on parameters like redshift or mass is a way to include a more complicated selection function without changing the luminosity integrals. 
$L$ and $\hat{L}$ are ``real'' and observed luminosities, respectively. $\sigma_{\hat{L}}$ is the uncertainty (model) of the measured luminosities, which could in principle depend on flux or solid angle.
$L_\mathrm{LM}$ is the luminosity coming from the $L-M$ relation and $\sigma_\mathrm{LM}$ is the scatter of the mass-luminosity function that is being used, which can also be variable during the cosmological fit. 
$P_I$ is the selection function and in the most simple case considered here it can be identified with a heavyside step function (1 for clusters above the flux limit and 0 below), so here it will just depend on $\hat{L}$ and $z$.
$\mathcal{N}(x,y,z)$ denotes the normal distribution probability density function at $x-y$ and with a standard deviation $z$. 
The probability function $p$ in Eq. \ref{eq:likelihood1} can be identified by
\begin{equation}
p = \frac{\langle N_\mathrm{det}\rangle}{\langle N \rangle}~,
\end{equation}
where $P_{i,\,\mathrm{det}}$ is the probability to detect a cluster with certain \textit{observed} quantities ($\hat{M},\hat{L},\hat{z},$...),
\begin{equation}
\label{eq:likelihood3}
\tilde{P}_{i} =  \frac{1}{\langle N_\mathrm{det} \rangle} \int \mathrm{d}z\,  \delta(\hat{z}_i) \cdot  \frac{\mathrm{d}V}{\mathrm{d}z}  \int \mathrm{d}M \, \Omega_\mathrm{frac}(z,M) \cdot \frac{\mathrm{d}n}{\mathrm{d}M} \int \mathrm{d}L\, \mathcal{N}(L,\hat{L}_i,\sigma_{\hat{L}_i}) \cdot \mathcal{N}(M,\hat{M}_i,\sigma_{\hat{M}_i}) \cdot \mathcal{N}(L,L_\mathrm{LM},\sigma_\mathrm{LM}) \cdot P_I~,
\end{equation}
where the redshift is assumed to be perfectly known (modeled by a delta function). $\hat{M}_i$ and $\sigma_{\hat{M}_i}$ are the measured total mass and its standard deviation, respectively.
The probability for the missed sources is simply
\begin{equation}
(1-p) = \frac{\langle N_\mathrm{mis} \rangle}{\langle N \rangle}~.
\end{equation}

Putting all these derivations together one simplifies Eq. \ref{eq:likelihood1} to,
\begin{equation}
\mathcal{L} = \underbrace{\frac{\langle N \rangle^N}{\langle N \rangle^{N_\mathrm{det}} \langle N \rangle^{N_\mathrm{mis}}} }_{\mathrm{1}} \underbrace{\frac{1}{N_\mathrm{det}!}}_{\mathrm{constant}} \underbrace{ \frac{\langle N_\mathrm{mis} \rangle^{N_\mathrm{mis}}\,e^{- \langle N_\mathrm{mis} \rangle} }{N_\mathrm{mis}!} }_{\mathrm{for}\,N_\mathrm{mis}\in[0,\infty] = 1} \cdot e^{- \langle N_\mathrm{det} \rangle} \prod \limits_{i=1}^{N_\mathrm{det}} \langle N_\mathrm{det} \rangle \cdot \tilde{P_i}  ~.
\end{equation}
The third term is a Poisson likelihood which is equal to unity when marginalizing over $N_\mathrm{mis}$ from 0 to $\infty$.
As indicated, only the last term depends on model parameters which gives the likelihood as in \citet{2010MNRAS.406.1759M,Mantz2015},
\begin{equation}
\label{eq:likelihood2}
\mathcal{L} \propto  e^{- \langle N_\mathrm{det} \rangle} \prod \limits_{i=1}^{N_\mathrm{det}} \langle N_\mathrm{det} \rangle \cdot \tilde{P}_{i}~.
\end{equation}

\section{Gas mass fraction}
\label{ap:fgas}
\begin{table}
	\renewcommand{\arraystretch}{1.0}
	\footnotesize
	\centering
	\begin{tabular}{ccc}
	\hline
	Cluster & $f_\mathrm{gas, 2500}$ & $\Delta f_\mathrm{gas, 2500}$ \\
	\hline
	2A0335 & 0.107 & 0.002 \\
	A0085 & 0.087 & 0.001 \\
	A0119 & 0.068 & 0.005 \\
	A0133 & 0.078 & 0.005 \\
	A0262 & 0.065 & 0.001 \\
	A0399 & 0.126 & 0.004 \\
	A0400 & 0.051 & 0.003 \\
	A0401 & 0.107 & 0.002 \\
	A0478 & 0.092 & 0.007 \\
	A0496 & 0.076 & 0.003 \\
	A0576 & 0.059 & 0.005 \\
	A0754 & 0.027 & 0.003 \\
	A1060 & 0.046 & 0.002 \\
	A1367 & 0.063 & 0.003 \\
	A1644 & 0.078 & 0.003 \\
	A1650 & 0.076 & 0.001 \\
	A1651 & 0.093 & 0.004 \\
	A1656 & 0.085 & 0.002 \\
	A1736 & 0.093 & 0.005 \\
	A1795 & 0.088 & 0.002 \\
	A2029 & 0.095 & 0.002 \\
	A2052 & 0.111 & 0.001 \\
	A2063 & 0.065 & 0.003 \\
	A2065 & 0.087 & 0.002 \\
	A2142 & 0.080 & 0.004 \\
	A2147 & 0.104 & 0.008 \\
	A2163 & 0.133 & 0.004 \\
	A2199 & 0.075 & 0.001 \\
	A2204 & 0.090 & 0.004 \\
	A2244 & 0.095 & 0.002 \\
	A2255 & 0.085 & 0.004 \\
	A2256 & 0.093 & 0.001 \\
	A2589 & 0.065 & 0.002 \\
	A2597 & 0.097 & 0.001 \\
	A2634 & 0.045 & 0.002 \\
	A2657 & 0.077 & 0.005 \\
	A3112 & 0.070 & 0.004 \\
	A3158 & 0.093 & 0.002 \\
	A3266 & 0.087 & 0.007 \\
	A3376 & 0.046 & 0.002 \\
	A3391 & 0.066 & 0.004 \\
	A3395 & 0.055 & 0.002 \\
	A3526 & 0.067 & 0.001 \\
	A3558 & 0.069 & 0.004 \\
	A3562 & 0.066 & 0.004 \\
	A3571 & 0.079 & 0.002 \\
	A3581 & 0.058 & 0.003 \\
	A3667 & 0.094 & 0.002 \\
	A4038 & 0.077 & 0.006 \\
	A4059 & 0.068 & 0.001 \\
	EXO0422 & 0.075 & 0.004 \\
	HydraA & 0.075 & 0.003 \\
	IIIZw54 & 0.078 & 0.002 \\
	MKW3S & 0.071 & 0.001 \\
	MKW4 & 0.044 & 0.001 \\
	MKW8 & 0.048 & 0.003 \\
	NGC1399 & 0.041 & 0.001 \\
	NGC1550 & 0.052 & 0.001 \\
	NGC4636 & 0.020 & 0.000 \\
	NGC5044 & 0.043 & 0.000 \\
	NGC507 & 0.044 & 0.002 \\
	RXCJ1504 & 0.109 & 0.008 \\
	S1101 & 0.077 & 0.002 \\
	ZwCl1215 & 0.081 & 0.004 \\
	\hline
	\hline
\end{tabular}
\caption{Gas mass fraction values for the clusters at $R = \Delta_{2500}$, as used in Section \ref{ch:gasmass_result}.}
\label{tab:fgas}
\end{table}


\bsp	
\label{lastpage}
\end{document}